\documentclass[aps,twocolumn,floatfix,sort,superscriptaddress]{revtex4}

\usepackage[latin9]{inputenc}
\usepackage[T1]{fontenc}
\usepackage[english]{babel}
\usepackage{amsmath,amsfonts}
\usepackage{color}
\usepackage{textcomp}

\usepackage[section]{placeins}

\usepackage{graphicx}
\usepackage{setspace}
\usepackage{subscript}
\usepackage[unicode=true]{hyperref}

\usepackage{epstopdf}

\makeatletter
\pdfpageheight\paperheight
\pdfpagewidth\paperwidth  
\@ifundefined{textcolor}{}
{
 \definecolor{BLACK}{gray}{0}
 \definecolor{WHITE}{gray}{1}
 \definecolor{RED}{rgb}{1,0,0}
 \definecolor{GREEN}{rgb}{0,1,0}
 \definecolor{BLUE}{rgb}{0,0,1}
 \definecolor{CYAN}{cmyk}{1,0,0,0}
 \definecolor{MAGENTA}{cmyk}{0,1,0,0}
 \definecolor{YELLOW}{cmyk}{0,0,1,0}
}

\setcitestyle{super}

\makeatother

\begin{document}

\title{Real-space collapse of a polariton condensate}

\author{L.~Dominici}
\affiliation{NANOTEC, Istituto di Nanotecnologia--CNR, Via Arnesano, 73100 Lecce, Italy}
\affiliation{Istituto Italiano di Tecnologia, IIT-Lecce, Via Barsanti, 73010 Lecce, Italy}
\email{lorenzo.dominici@gmail.com}
\author{M.~Petrov}
\affiliation{SOLAB, Spin Optics Lab, St.~Petersburg State University, 198504 St.~Petersburg, Russia}
\author{M.~Matuszewski}
\affiliation{Institute of Physics, Polish Academy of Sciences, Al.~Lotnikow 32/46, 02-668 Warsaw, Poland}
\author{D.~Ballarini}
\affiliation{NANOTEC, Istituto di Nanotecnologia--CNR, Via Arnesano, 73100 Lecce, Italy}
\author{M.~De~Giorgi}
\affiliation{NANOTEC, Istituto di Nanotecnologia--CNR, Via Arnesano, 73100 Lecce, Italy}
\author{D.~Colas}
\affiliation{IFIMAC, F\'{i}sica Teorica de la Materia Condensada, UAM, 28049 Madrid, Spain}
\author{E.~Cancellieri}
\affiliation{Laboratoire Kastler Brossel, UPMC-Paris 6, \'{E}NS et CNRS, 75005 Paris, France}
\author{B.~Silva~Fern\'{a}ndez}
\affiliation{NANOTEC, Istituto di Nanotecnologia--CNR, Via Arnesano, 73100 Lecce, Italy}
\affiliation{IFIMAC, F\'{i}sica Teorica de la Materia Condensada, UAM, 28049 Madrid, Spain}
\author{A.~Bramati}
\affiliation{Laboratoire Kastler Brossel, UPMC-Paris 6, \'{E}NS et CNRS, 75005 Paris, France}
\author{G.~Gigli}
\affiliation{NANOTEC, Istituto di Nanotecnologia--CNR, Via Arnesano, 73100 Lecce, Italy}
\affiliation{Universit\'{a} del Salento, Via Arnesano, 73100 Lecce, Italy}
\author{A.~Kavokin}
\affiliation{CNR-SPIN, Tor Vergata, viale del Politechnico 1, I-00133 Rome, Italy}
\affiliation{Russian Quantum Center, 143025 Skolkovo, Moscow Region, Russia}\author{F.~Laussy}
\affiliation{IFIMAC, F\'{i}sica Teorica de la Materia Condensada, UAM, 28049 Madrid, Spain}
\affiliation{Russian Quantum Center, 143025 Skolkovo, Moscow Region, Russia}
\author{D.~Sanvitto}
\affiliation{NANOTEC, Istituto di Nanotecnologia--CNR, Via Arnesano, 73100 Lecce, Italy}

\begin{abstract}
\begin{singlespace}
  \textbf{Polaritons in microcavities are versatile quasi-2D bosonic
    particles with a high degree of coherence and strong
    nonlinearities, thanks to their hybrid light-matter
    character~\citep{Amo2009}. In their condensed
    form~\citep{Kasprzak2006}, they display striking quantum
    hydrodynamic features analogous to atomic Bose-Einstein
    condensates, such as long-range order coherence,
    superfluidity~\citep{Amo2009a} and quantized
    vorticity~\citep{Nardin2011a}.  Their variegated dispersive and
    dissipative properties~\citep{Luk2013,Vladimirova2010}, however,
    set significant differences from their atomic
    counterpart~\citep{Carusotto2013}. In this work, we report the
    unique phenomenology that is observed when a pulse of light
    impacts the polariton vacuum: the condensate that is
    instantaneously formed does not splash in real space but instead
    coheres into an enigmatic structure, featuring concentric rings
    and, most notably, a sharp and bright peak at the center.  Using a
    state-of-the-art ultrafast imaging with 50~fs time steps, we are
    able to track the dynamics of the polariton mean-field
    wavefunction in both real and reciprocal space.  The observation
    of the real-space collapse of the condensate into an extremely
    localized---resolution limited---peak is at odd with the repulsive
    interactions of polaritons and their positive effective mass. An
    unconventional mechanism is therefore at play to account for our
    observations. Our modeling suggests that self-trapping due to a
    local heating of the crystal lattice---that can be described as a
    collective polaron formed by a polariton condensate---could be
    involved. These observations hint at the fascinating fluid
    dynamics of polaritons in conditions of extreme intensities and
    ultrafast times.  } \end{singlespace}
\end{abstract}

\maketitle

Microcavity polaritons have been praised for their fast response
times~\citep{Sich2012,dominici2014} and ease of manipulation as well
as detection, inherited by the photonic component, while keeping a
strong nonlinear character~\citep{Ballarini2013}, conferred by the
excitons~\citep{Vladimirova2010}. This makes them increasingly strong
contenders in the field of interacting quantum fluids, where they have
demonstrated the prevailing phases of strongly correlated
systems~\citep{Amo2009}, including Bose--Einstein
condensation~\citep{Kasprzak2006}, superfluidity~\citep{Amo2009a},
e.g., scatterless flow, quantized
vortices~\citep{Sanvitto2010,Nardin2011a}, together with rich
spinorial patterns~\citep{Manni2013} and nonlinear interference
effects. Polaritons have also demonstrated their suitability to
investigate another mainstream concept fueled by dispersive and
dissipative nonlinearities: shock waves and solitons, respectively
characterized by step disturbances moving in the medium with sound
velocity and by self-localization in space or shape preservation in
time. Beyond polariton
fluids~\citep{Sich2012,Grosso2012,Pigeon2011,Sich2014,Tanese2013,Zhang2010},
these have also drawn much attention in nonlinear
media~\citep{Ghofraniha2012,Wan2006}, atomic Bose-Einstein condensates
(BECs)~\citep{Chang2008,Dutton2001,Eiermann2004,Hai2009,Joseph2011,Kamchatnov2004,Marchant2013,Simula2005}
or microcavities in general~\citep{Barland2002,Barland2012}. For
instance, the response of a nonlinear medium or atomic BECs to an
impinging blast resulting in the irradiation of shock
waves~\citep{Dutton2001,Barsi2007}, or the appearance of solitonic
states~\citep{Sich2012} have been recently reported.  Time-resolved
exploration of such effects in strongly correlated gases remains
largely unexplored and with the ultrafast imaging techniques now
available, one is at the dawn of a new era for the investigation of
quantum fluid dynamics.

While they have reproduced most of the known phenomenology of quantum
gases, polaritons also come with peculiarities of their own, such as
their dispersion relation or their short lifetime, making them
intrinsically out-of-equilibrium. Here, we report what appears to be a
unique phenomenology of these systems, observed after the sudden
coherent generation of a polariton condensate. The fluid undergoes a
space redistribution leading to a central localization of a great
number of polaritons, despite their repulsive interactions. The peak
that is formed reaches a localization ($\le 2\mu $m, resolution
limited) at least ten times sharper than the initial gaussian spot
injected by the laser (18.5 $\mu $m). It also gathers a large number of
particles, with a local enhancement up to 10 times the original
density of polaritons. This striking dynamics takes place in a few
ps. All these features can be tuned continuously with the excitation
power. Another interesting feature of the dynamics is the generation
of a shock wave at early times and concentric rings at later
times. Similar rings and shock waves have been observed in nonlinear
defocusing optical media or repulsive atomic
BECs~\citep{Kamchatnov2004,Wan2006,Barsi2007}, but the presence of a
central localised and enhanced peak has never been reported so far in
any system, to the best of our knowledge.

The sample we used consists of a 2$\lambda $ microcavity (MC)
containing a triple quantum well (QW) positioned in the maxima of the
electromagnetic field~\citep{Ballarini2013,Amo2009a}. An excellent
quality factor (photonic $Q=14000$) is given by two high reflectance
multilayers mirrors (DBR, distributed Bragg reflectors) embedding the
MC.  The strong coupling between the QW excitons and MC photons
manifests itself as an anticrossing of their original modes splitting
into two new normal modes, known as polaritonic upper (UPB) and lower
(LPB) branches (see supplementary figure S5). All the experiments shown
here are performed at zero detuning between the MC and the QW exciton,
both resonant at $\sim 836$ nm, and in a region of the sample clean
from defects in order to avoid any effect due to spatial
inhomogeneities. The device is kept at a temperature of 10~K. We
implemented an unltrafast imaging technique based on the off-axis
digital
holography~\citep{dominici2014,Anton2012,Nardin2011a,Schnars_book2005a}
to study the dynamics of the polariton flow with spatial and temporal
steps of 0.16~$\mu m $ and 50~fs respectively. A 130~fs or 3.5~ps
laser pulse, on resonance with the bottom of the LPB and circularly or
linearly polarized, is directed onto the sample at normal incidence
and the evolution of the polariton state in time is recorded by making
interfere the sample emission with a delayed reference beam into a CCD
camera. Digital elaboration in the reciprocal space of the
interferograms allows to retrieve the complex wavefunction of the
photonic emission, which is coherent with the polaritonic
wavefunction, and thus allows us to image both the amplitude and phase
of the light-matter fluid.  For energy resolved images a standard
550~cm long spectrometer is used before the CCD and the spectra are
time integrated. Additional details on the sample and the technique
can be found in Ref.~\citep{dominici2014} and the supplementary
material therein.

\begin{figure}[htbp]
\centering\includegraphics[width=\linewidth]{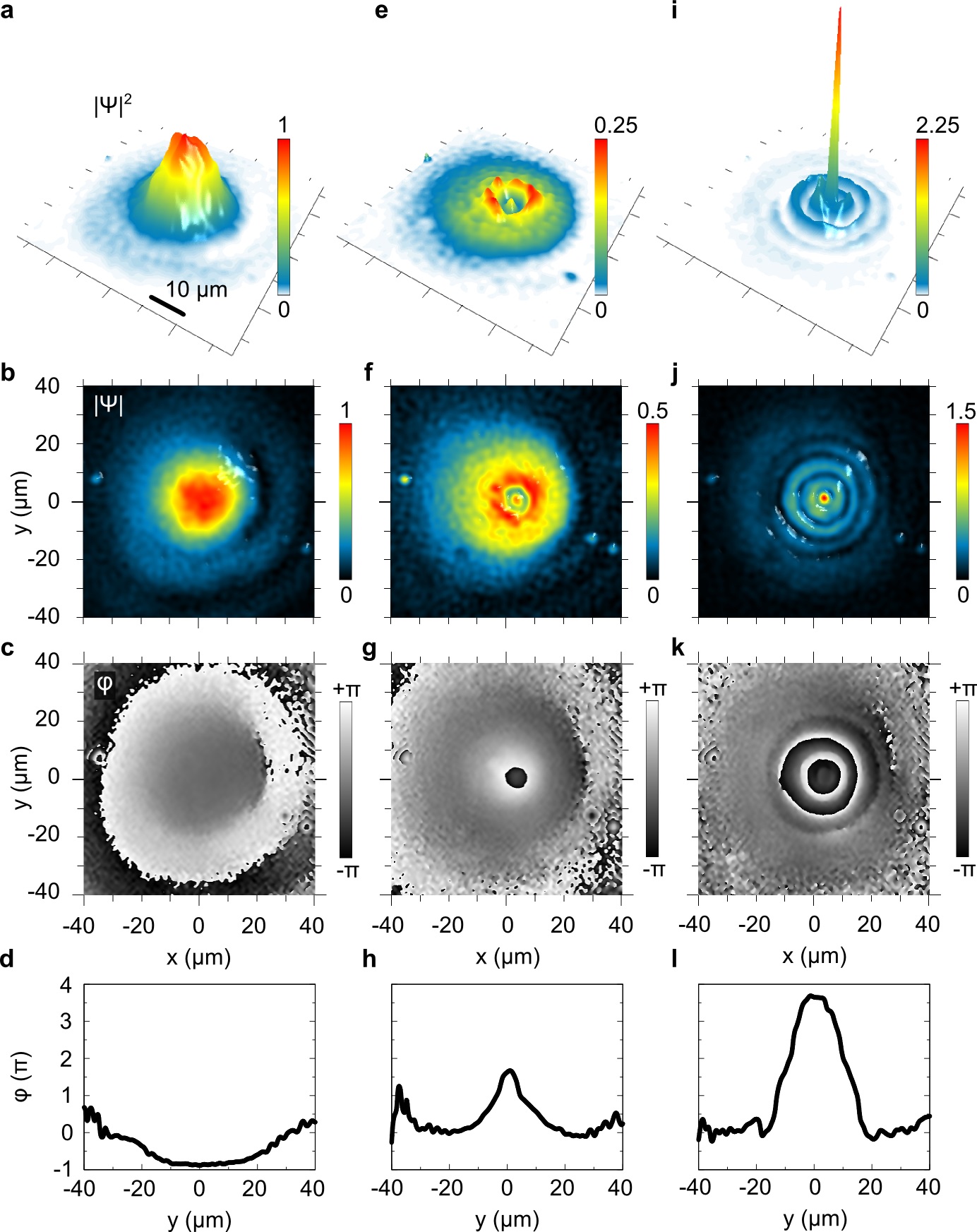}
\caption{\textbf{Snapshots of the polariton fluid density and phase at
    three significative instants.} \textbf{a,e,i}, (first row) Density
  maps of the planar polariton condensate on a
  $80\times80\protect\mu$m area as 3D view and \textbf{b,f,j} (second
  row) amplitude maps as 2D view. The three columns represent time
  frames at $t=0$~ps (a-d), 2.8~ps (e-h) and 10.4~ps (i-l).  These
  time frames correspond respectively to the pulse arrival, the
  ignition of the dynamical peak and its long-lived state sitting at
  the center of a ring structure (see also video
  S1--S3). \textbf{c,g,k} (third row) phase maps and \textbf{d,h,l}
  (fourth row) unwrapped phase profiles along the radius. The phase
  gradient subtends the superflow and here exhibits a reversal of the
  phase curvature, leading to the development of an opposite flow,
  toward the center. The total number of particles intially excited in
  the whole area is $250\times10^3$ polaritons as evaluated by an
  independent measure of the bare emission. }
\end{figure}

\begin{figure*}[htbp]
\centering\includegraphics[width=0.75\linewidth]{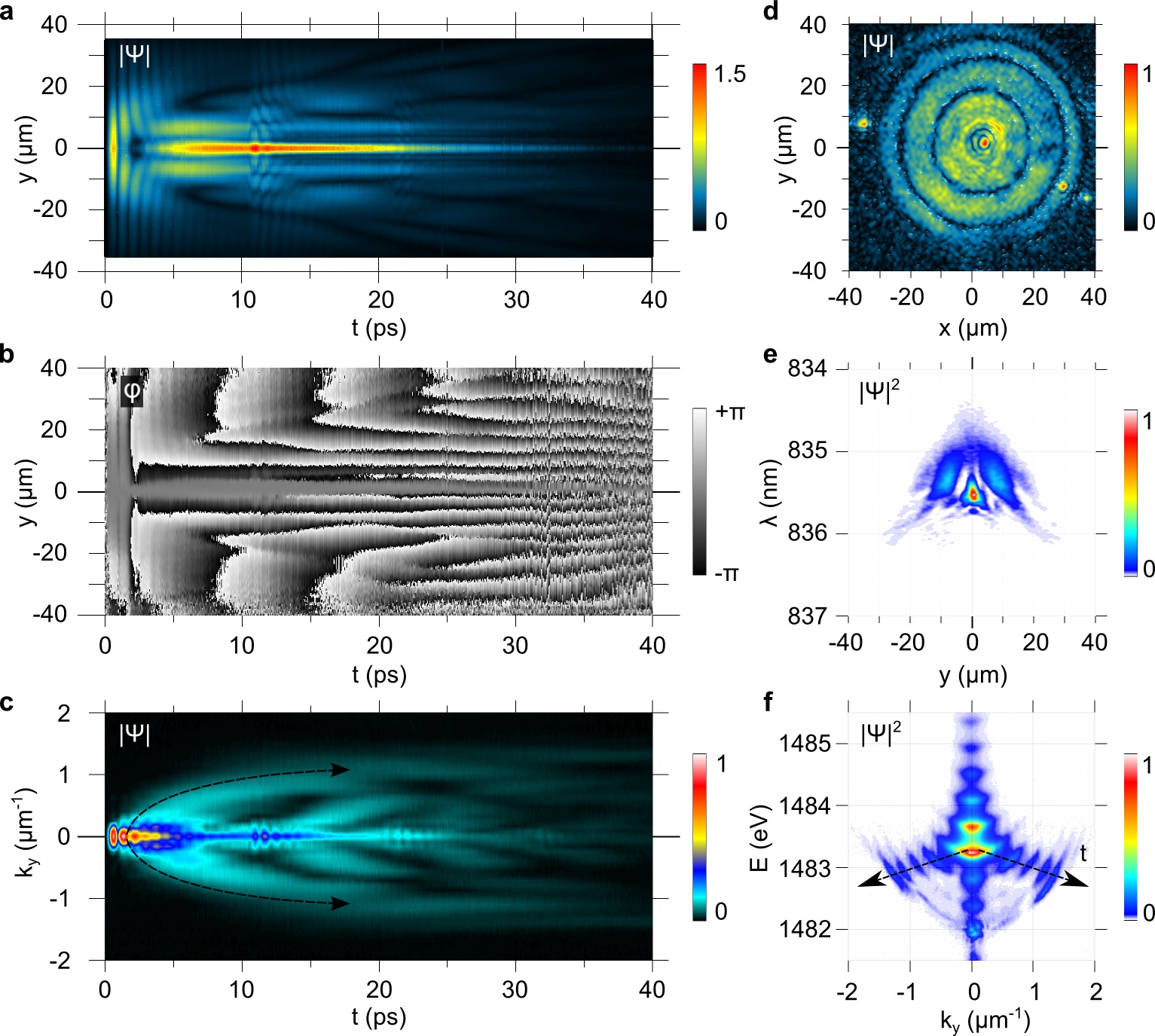}
\caption{\textbf{Dynamical charts of the complex wavefunction and
    amplitude maps.} \textbf{a}, Time-space chart of the polariton
  amplitude $|\Psi(t,r)|$ sampled with a timestep $\delta
  t=50$~fs. The polariton fluid oscillates with a Rabi period of about
  800~fs (vertical stripes in the map), while the central density
  rapidly decays to zero before starting to rise as a bright peak. An
  echo pulse due to a reflection from the substrate edge is visible at
  $t=11$~ps. \textbf{b}, Time-space chart of the phase $\Phi(t,y)$.
  Two diagonal horizon lines delimit an expanding region with large
  $\protect%
  \nabla\Phi$. \textbf{c}, The time evolution of amplitude in momentum
  space, $%
  |\Psi(t,k_{y})|$. The initial polariton population, featuring a very
  narrow $%
  \Delta k$ width (imparted by the photon packet), ejects an expanding
  disk developing into a ring. \textbf{d}, $|\Psi(x,y)|$ map at
  $t=26$~ps, showing the dark/bright ring structures. \textbf{e}, A
  $y-\lambda$ cut showing the energy of the fluid along the
  diameter. The central brightest spot is less blueshifted than its
  sides.  \textbf{f} Energy-momentum $E-k_{y}$ dispersion under the
  femtosecond coherent excitation. The dashed arrows depict the
  opening up in the $k$ space and are associated to the dashed lines
  in panel \textbf{c}.}
\end{figure*}

The dynamics of the polariton fluid, generated instantaneously by a
resonant laser pulse hitting the sample, is shown in
Fig.~1. Initially, the polariton distribution is simply a footprint of
the incoming laser spot, i.e., a gaussian of 18.5~$\mu $m
(FWHM). Panels (a-c) shown the sample emission at the pulse arrival,
in density (a), amplitude (b) and phase (c). The unwrapped phase along
a central diameter is shown in panel (d), with the shallow positive
curvature associated to a weak outward gradient.  The central density
of polaritons is initially 550~$\mu m^{-2}$. For the first few ps, the
fluid only slightly decreases its total number of particles---due to
radiative losses---however, at $\sim 2$ ps, the maximum density
suffers a sudden depletion, down to only a quarter of the initial
density. This is shown in panels (e,f) for the intensity and (g,h) for
the phase. This unanticipated jolt marks the beginning of the
redistribution of the fluid.  Note that the phase space profile starts
to manifest a negative curvature with a reversal point of the phase
gradient ($\nabla \phi $) at $r=20\mu $m. Such an inversion in the
phase gradient is an indication of a change in the fluid direction,
from waves expanding outwards, to contracting towards the centre. This
behaviour is then followed by the appearance of the bright and sharp
central peak, contoured by concentric rings (i,j). This central peak
collects 6\% of the total particles still present in the fluid, and is
surprisingly brighter than the initial spot was in the same area (here
by a factor of two, up to ten times in other realizations, see
supplementary Fig.~S3) despite the total population having largely
decreased (to less than $1/3$ after 10~ps). Its localization goes
below the resolution of our experimental setup and is thus less than
2~$\mu $m in width.  The peak is furthermore extremely robust as it
occurs over the whole sample area, at different MC/QW detunings and
even sustains motion, as it actually propagates if imparted with an
initial momentum (see Supplementary Material).  It is observed in the
time-integrated camera images of the direct emission before any
subsequent digital elaboration, which excludes any artifact of the
technique (see also Fig. S3 in the Supplementary Material).

The full dynamics of the polariton fluid and the connection between
the density accumulation and the radial flows are further studied in
Fig.~2. Here are shown the amplitude and phase profiles versus time
{[}panels (a) and (b){]}. The amplitude chart reveals the bright peak
as a central horizontal line which reaches its maximum intensity at a
time of $\sim$11 ps. The vertical stripes observed in the first 5-6~ps
are Rabi oscillations between the excitonic and photonic fields. The
period of $T{}_{R}\approx800$~fs corresponds to the energy separation
between the UPB and LPB of 5.4 meV, while their fast decay is due to
the rapid scattering of the UPB polaritons. The Rabi oscillations are
triggered by the femtosecond pulse that excites simultaneously both
polariton branches (9 nm energy width, see also supplementary Fig.~S1
and Ref.~\citep{dominici2014}). They do not, however, play an
important role in the observed phenomenology, since the dynamical
localization also happens under the excitation of the LPB alone by a
picosecond laser pulse (0.35 nm width, see supplementary Fig.~S2). This
confirms the robustness of this peak.  Since it has an homogeneous
phase, as can be seen from the phase graph in Fig. 2(b), it is a
standing wave. We also note that a neat time-space cone marks a
boundary between two regions: an expanding internal domain with almost
horizontal black and white (b/w) bands (strong inward $\nabla\Phi$)
and an external domain with almost vertical b/w bands (null or weak
$\nabla\Phi$).  This is a clear evidence that a circular front of
phase disturbance (marked also by a low density) is expanding
($\sim1\mu$m/ps), leaving after its passage a fragmentation into
multiple rings, as illustrated in the still image of panel~(d). If an
attractive term is at play here, it thus seems to be in the course of
expanding its range of action. To further characterize the flow
dynamics and its role in the formation of the central peak, we extend
our study to the reciprocal space, reporting the 1D cross section of
the $k_{x},k_{y}$ plane in panel (c) of figure 2. The initial $\Delta
k$ width of the polariton population created by the laser pulse is
very small ($\sim0.24\mu m^{-1}$ FWHM) and concentrated around
$k=0$. After a couple of picoseconds, the fluid suddenly ejects a disk
in $k$-space which stabilizes within 10~ps into a ring at finite
momenta around $|k|=1\mu m^{-1}$. These are clearly associated to the
inner growing flow. Looking at the onset of the Rabi oscillations
allows us to make a correspondence between real and reciprocal space,
indicating that the central bright peak in real space is associated to
the finite momenta travelling waves in reciprocal space. The energy of
these waves is shown in the integrated dispersion of
Fig. 2(f)~\citep{nota1}. Note that, as oberved in the dispersion
plotted in Fig. 2(f), the ring in $k$ space gradually expands from
$k=0$ at early times (maximum blushift of the polariton pupulation) to
the value of $|k|=1.25\mu m^{-1}$ when the dispersion is redshifted to
its original bare energy. The space-energy profile in figure 2(e)
explains in part the inner flow. Here it is clear that the central
part of the spot is redshifted with respect to the sides, hinting at
the presence of an effective attractive potential responsible for the
central peak of high polariton density. Although it is not clear at
the outset what the origin of such a potential is, it is consistent
with an attractive term (nonlinearity inversion) or with nonlocal
interactions, i.e., $k$-dependent
blueshift~\citep{Baumberg2005a,Luk2013}. In our case, we may
reasonably infer that the inward coherent waves generated by such a
potential interfere in the center with opposite $k_{r}$ vectors,
explaining the central accumulation enhanced by a polar scaling of
$1/r$ which is typical of an interfering ring wave (see figure S7 in
the SM).

\begin{figure}[tb]
\centering\includegraphics[width=\linewidth]{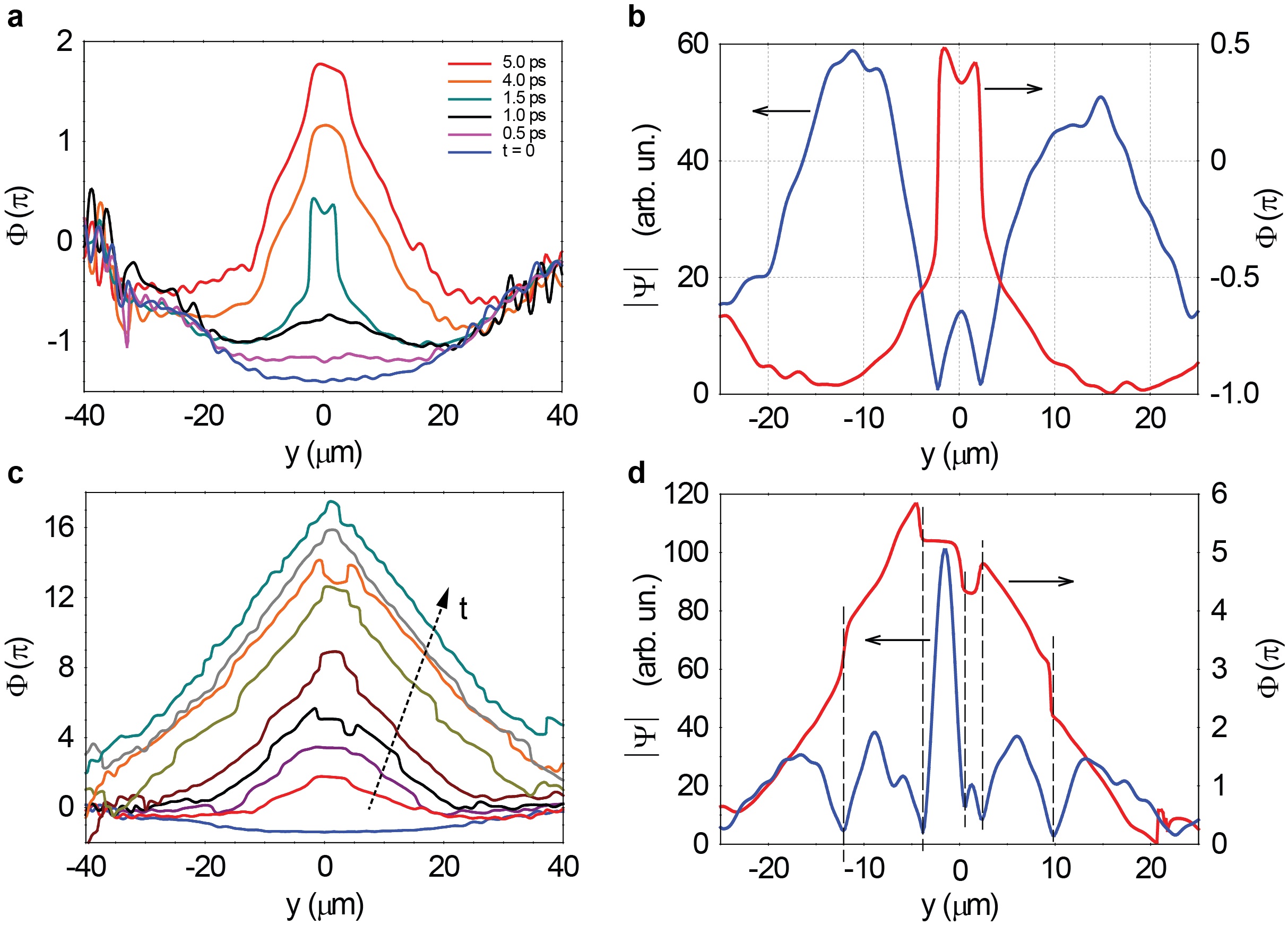}
\caption{\textbf{Phase crosscuts during the fluid evolution and
    signature of dark ring solitons.} \textbf{a}, Unwrapped radial
  phase profile at early time, showing the reversal of the phase
  curvature. \textbf{b}, The sudden phase switch at $t=1.5$~ps is
  shown together with the associated amplitude profile. The dip in the
  intensity with the $\protect\pi$-jump in phase is a signature of
  radial interference and of a possible dark ring soliton, surrounding
  the bright peak. \textbf{c}, Radial phase profile at later time,
  taken each 5~ps over a 0--40~ps timespan. The phase slope increases
  with time up to $\nabla\phi\sim2\pi/5\mu m\simeq1.25\mu
  m^{-1}$. \textbf{d}, Amplitude and phase profiles at 16~ps showing
  that a nonlinear interference is reshaping the fluid in a series of
  concentric rings.}
\end{figure}

To verify the nature of the interference, we report the phase and
amplitude profiles at different time stills. In Fig.~3(a), the phase
profile during the first 5~ps is shown. Soon after the phase reversal,
a sudden, almost instantaneous switch happens at $t=1.5$~ps. The rigid
switch is a $\pi$-jump, corresponding to the dark notch in the
density, as seen in panel (b). This is a typical sign of an
interference between counterpropagating waves. In the following
dynamics, as already said, the reversed gradient tends to grow and
expand all over the originally excited area, as depicted in panel (c)
with profiles at 5 ps intervals. Other $\pi$-jumps in the phase can be
observed at a later time, corresponding to dark rings in the density,
see Fig.~3(d). This is an evidence that interference phenomena of
coherent waves are acting in reshaping the fluid density as a series
of concentric rings.  It also suggests a possible link to---without
being a ``per-se'' proof of---ring dark solitons
(RDS)~\citep{Kivshar1994}. These specific solutions to nonlinear
Schr\"odinger equations (NLSE) under repulsive interactions are
predicted to become stable in the case of 2D fluids such as
polaritons~\citep{Rodrigues2014}. It appears that a RDS indeed holds
for the first dark ring around the rising bright peak, given its
stability for several tens of ps.

\begin{figure*}[htbp]
\centering\includegraphics[width=.8\linewidth]{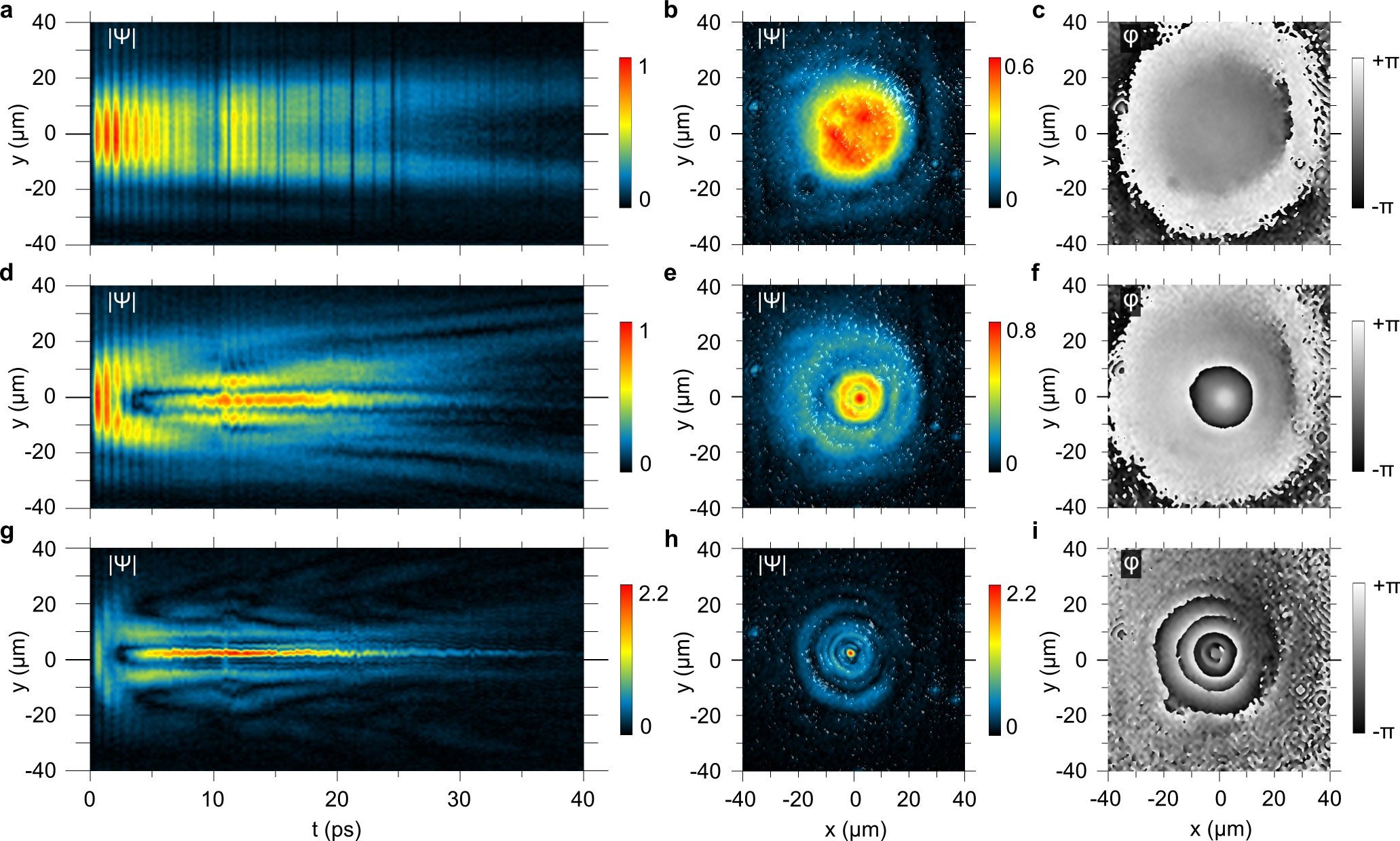}
\caption{\textbf{Time-space charts and space maps for different
    density regimes of excitation.} Time evolution of the radial
  modulus $|\Psi(t,y)|$ for three different powers (\textbf{a,d,g},
  left column), and relative amplitude $|\Psi(x,y)|$ (\textbf{b,e,h})
  and phase $\Phi(x,y)$ (\textbf{c,f,i%
  }) maps at $t=12$~ps (mid and right columns,
  respectively). Increasing the initial density leads to a faster
  central depletion and stronger rise-back reaction. In the third row
  the dominating feature is the bright peak, with an enhancement
  factor of almost 5 in intensity, while outradiated waves are faster
  but almost canceled out on a relative scale. This demonstrates the
  strong nonlinearities acting in the central gathering of polaritons
  and in setting the radial $k$ and speed of the ring waves. The three
  rows refer to initial total populations (top density) of
  $25\times10^3$ (55$\mu m^{-2}$), $125\times10^3$ (275$\mu m^{-2}$)
  and $450\times10^3$ polaritons (1000$\mu m^{-2} $), respectively.}
\end{figure*}

The nonlinear nature of the effect is demonstrated in Fig.~4. At low
density, Fig.~4(a,b,c), the condensate behaves as expected from any
fluid freely released, with a small diffusion and remaining
homogeneous in both density and phase, as well as, in our case, Rabi
oscillations at earlier times. It is also clear that the expansion
speed imparted by the small initial $\Delta k$ is negligible. At
5-times higher excitation power {[}Fig. 4(d,e,f){]}, there is a
density redistribution of polaritons to form the central localization
peak, surrounded by ring structures and out-radiating shock waves,
however, yet without a strong enhancement and a moderate phase
reversal. In Fig.4(g,h,i), at 18 times the initial pumping power, the
structure gets fully formed, with a central peak gathering over twice
the population locally present at the initial time and with a much
steeper phase reversal, giving rise to the striking structure in
Fig.~4h. Here again we emphasize that the central peak is resolution
limited and is likely sharper than is resolved in our
experiment. Additional examples are provided in the supplementary
Fig.~S3 for an extended set of excitation densities.

We now discuss which physical mechanism could be responsible for such
a remarkable phenomenology. The strong-coupling regime of light and
matter at the core of the polariton physics can lead to distinctive
dispersive and dissipative
nonlinearities~\citep{Takemura2014,Vladimirova2010,Weiss2007,Schaefer1997}.
For instance, polaritons support dissipative
solitons~\citep{Ostrovskaya2012}, have demonstrated bistability
domains with a switching on/off of both bright and dark
solitons~\citep{Larionova2008} as well as moving bright solitons along
a steadily pumped background~\citep{Sich2012,Sich2014} (this last
based on the negative curvature of the polariton dispersion above the
inflexion point). All these features are accountable by one of the
several models used to describe polariton fluids. While the positive
nonlinearities intrinsic to polariton interactions, due to the
excitonic repulsions, are supposedly able to force the expansion and
reshaping of a polariton fluid and to sustain dark
solitons~\citep{Kivshar1994}, possibly shedding light to some aspect
of our experiment, there is no documented mechanism to explain the
most striking feature: the real-space collapse in the center of the
spot. Real space localization could in principle appear under
negative, i.e., attractive, nonlinearities~\citep{Vladimirova2010}. As
we review below all the obvious candidates to account for the observed
phenomenology, we can rule out this and other tempting explanations.
Our analysis will show that the most likely origin of the nonlinear
activation of a central bright spot is the self-trapping of a
polariton condensate by a type of collective polaron effect.

\begin{figure*}[tb]
\centering\includegraphics[width=.8\linewidth]{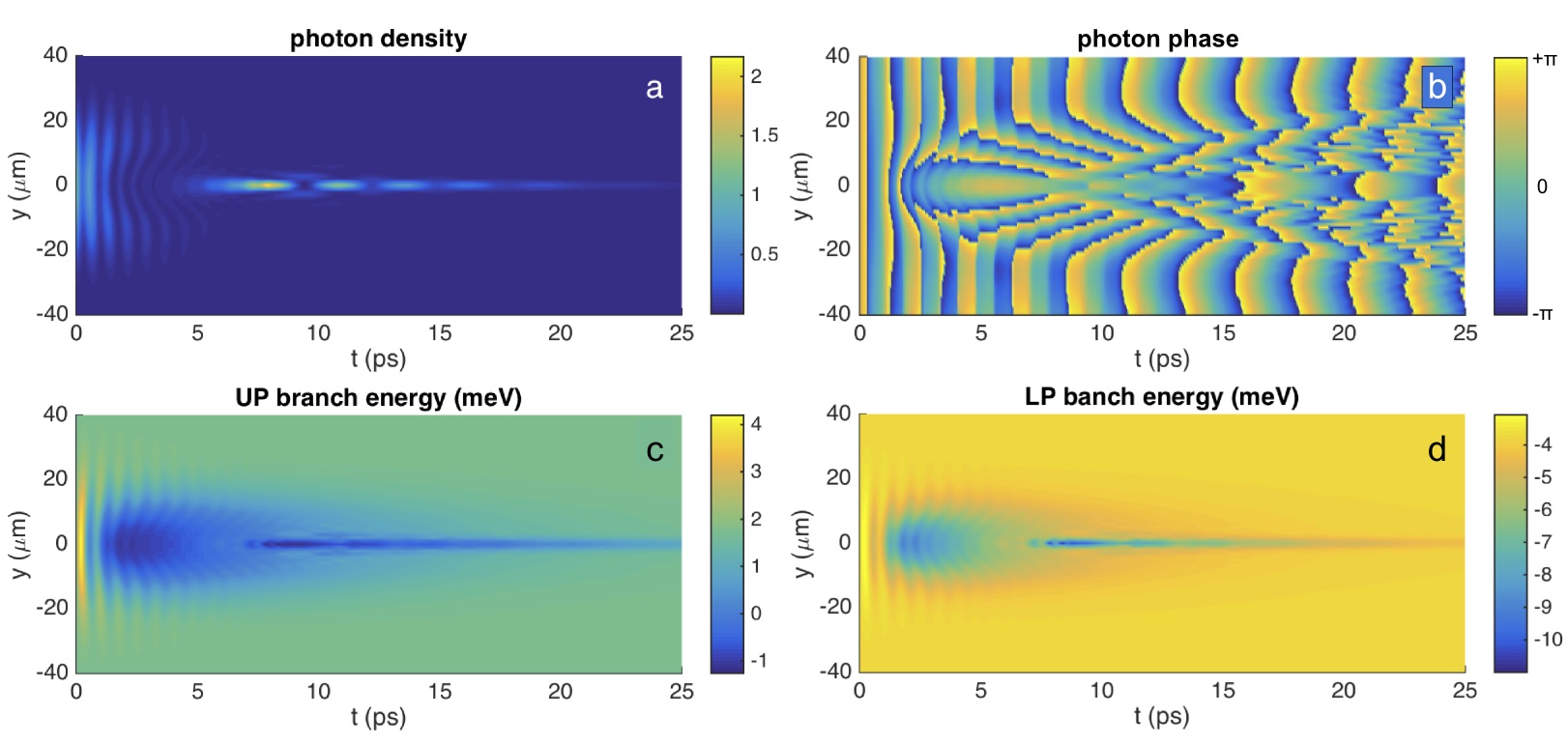}
\caption{The calculated magnitude of the polariton condensate as a
  function of y-coordinate and time (a) together with its phase (b).
  The corresponding calculated energy profiles for the upper and lower
  polariton branches are shown in panels (c) and (d), respectively.}
\end{figure*}

A first possibility to explain the self-localization is for the
polariton population at early times to undergo a transition to the
weak coupling regime (screening and reduction of coupling). Although
this can be partially happening at the largest power (given the
excited density there is $0.33\times10^{3}\mu m^{-2}$ per QW and
approaches values of the so-called Mott density, in the order of
$2\div5\times10^{3}\mu m^{-2}$), it does not explain the effect since
(i) the blueshift of the LPB is continuous and does not reach the
photonic mode; (ii) increasing power/density does not increase the
size of the bright peak, as expected if merely enlarging the size of
an above-threshold region; (iii) the total intensity decays with the
LP polariton lifetime of 10~ps (see supplementary Fig.~S1) and (iv)
transition to weak coupling for excitation below the band edge would
manifest as a higher blueshift as follows from the nonlinear
Kramers-Kronig relations~\cite{Garmire2000,Taranenko2005}, while we
observe a lesser blueshift in the centre with respect to the side
bright ring, as shown in Fig.~2(e). An exciton reservoir, separate
from the polariton condensate, is known to play a significant role in
many polariton experiments. We can exclude it in our case for the
following reasons: (i) the effect persists when resonantly exciting
the lower polariton branch, a configuration that does not populate the
reservoir (see the supplementary Fig.~S2 on the picosecond experiment),
(ii) a moving bright peak is observed when exciting with nonzero
initial in-plane wavevector~$k$, ruling out a reservoir that would
need to be dragged by the structure as it propagates, which is
impossible given the heavy mass of the reservoir excitons (see the
supplementary Fig.~S6 on the moving peak), (iii) the effect does not
show any strong polarization dependence. These considerations thus
exclude the case of a dissipative bright
soliton~\citep{Ostrovskaya2012} predicted under cw pumping where the
localization is sculpted by an interplay between source and decay
regions and the compensation of their steady flows.  The appealing
recourse to attractive interactions cannot be sustained either. While
polariton attractions are possible due to the various superexchange
processes through dark excitons or bi-exciton states, or due to Van
der Waals forces, none of these mechanisms can account for the
experiment in a careful analysis.  The former mechanism should be
strongly polarisation dependent~\citep{Vladimirova2010,Takemura2014},
which is not our case, and the latter are too small with respect to
the repulsive Coulomb or exchange
terms~\citep{Vladimirova2010}. Similarly, nonlocality ($k$-dependence,
which could also lead to a negative dispersion in the centre of the
spot~\citep{Baumberg2005a}) of one or more of these terms and even
retardation effects are negligible as well~\citep{Luk2013} and fail to
produce the real-space collapse in numerical
simulations. Intriguingly, the dynamical Casimir
effect~\citep{zhang2014} recently proposed~\citep{Koghee2014} in a
configuration very similar to our experiments, was predicted to
generate finite momentum excitations from the vacuum adding up to the
suddenly excited fluid at $k=0$.  However, this effect, was considered
only for the homogeneous 1D polariton at zero temperature, and the
model is not ready to be compared at its stage of development with the
configuration of our experiment.

The failures of these analyses point at an unconventional mechanism
ruling the high-density, ultrafast dynamics of polaritons. Given that
the object appears to be self-sustained, it is important to elucidate
its nature, as it may have important applications, especially as the
control of ultra-sharp localized light peaks is clearly of
technological interest, for instance for high-resolution displays or
memory units. One of the unconventional scenarios that we found to be
fairly consistent with most the hypotheses and observations of our
experiment involves a sort of collective polaron effect. The recent
work by Klembt~\textit{et al.}~\citep{Klembt2015} shows that the
resonant pumping of exciton polaritons into a microcavity may result
both in cooling or heating of the crystal lattice depending on the
initial lattice temperature and the optical pump power. In our
experiments, realised at a cryogenic temperature and sufficiently high
pump power, one should expect such a local heating of the crystal
lattice due to the polariton Auger process, followed by the emission
of a cascade of acoustic phonons. The probability of this process is
quadratic in the polariton density. The heating results in the local
band-gap renormalisation which is responsible for the red-shift of the
exciton energy. The heating by 20-30 degrees results in a redshift of
the exciton energy of 1-2 meV, which is sufficient for trapping the
polariton condensate. In this way, a trap in real space is formed
under the pump spot. It becomes deeper as more polaritons are getting
trapped, thus providing a positive feedback that stabilizes the
self-trapping process and explain the robustness of the effect. We
have modeled this mechanism through a generalised Gross-Pitaevskii
equation, described in the Supplementary Material. Figure~5 shows a
result of the numerical simulation for the wavefunction of the
polariton condensate ruled by this process. The upper panel shows the
real-space dynamics and next panels the energies of the low and upper
polariton branches, respectively. Beyond the dynamics of the Rabi
oscillations, in particular their bending, the model also reproduces
the self-localization in good qualitative agreement with the
experimental data.

In conclusions, we have observed the dynamical appearance of a bright
and sharp peak sitting at the center of a series of concentric rings
in a polariton fluid generated by the sudden excitation from a
resonant laser pulse at $k=0$. The peak that appears at high pumping
is robust to other variations in the experimental parameters
(detuning, momentum, etc.), is resolution limited and gathers up to
ten times the population initially present in its area. This striking
structure cannot be explained by any of the conventional mechanisms
such as loss of strong-coupling nor by the common models of polariton
dynamics, including Gross-Pitaevskii type of equations with or without
reservoirs and/or attractive interactions. We have provided a possible
intepretation in terms of the collective polaron effect, resulting in
a self-trapping of the polariton condensate. Our results show that
much is left to explore in the high-density and ultrafast dynamics of
polaritons, with a striking and unique phenomenology that could open
new areas of research and applications.


\begin{acknowledgments}
  We acknowledge R. Houdr\'{e} for the growth of the microcavity
  sample and the project ERC POLAFLOW for financial support. This work
  has been partially funded by the Quandyde project of the ANR France
  and by the CLERMONT4 Network Program. MM acknowledges support from
  the National Science Center grant DEC-2011/01/D/ST3/00482.
\end{acknowledgments}

\newpage
\onecolumngrid

\setcounter{equation}{0}
\setcounter{figure}{0}
\setcounter{table}{0}
\setcounter{page}{1}
\makeatletter
\renewcommand{\theequation}{S\arabic{equation}}
\renewcommand{\thefigure}{S\arabic{figure}}
\renewcommand{\bibnumfmt}[1]{[S#1]}
\renewcommand{\citenumfont}[1]{S#1}
\pagenumbering{roman}





\setcounter{MaxMatrixCols}{10}

\def\beq{\begin{equation}}
\def\eeq{\end{equation}}
\def\ba{\begin{align}}
\def\ea{\end{align}}
\newcommand{\er}{\mathbf{r}}
\newcommand{\ee}{{\rm e}}
\newcommand{\nR}{n_{\rm R}}
\newcommand{\gammac}{\gamma_{\rm C}}
\newcommand{\gammar}{\gamma_{\rm R}}
\newcommand{\gr}{g_{\rm R}}
\newcommand{\gc}{g_{\rm C}}
\newcommand{\tauQ}{tau_{\rm Q}}


\vfill\eject


\begin{center}
\textbf{\large Supplementary Material}
\end{center}

 

In this Supplementary Material, we describe the supplementary movies,
some supporting experimental figures and discuss several theoretical
models used in attempt to describe the collapse in real space of the
polariton condensate.

\section*{Supporting experimental figures S1-S7}

\vskip-0.6cm
\begin{figure}[tbhp]
\centering\includegraphics[width=.45\linewidth]{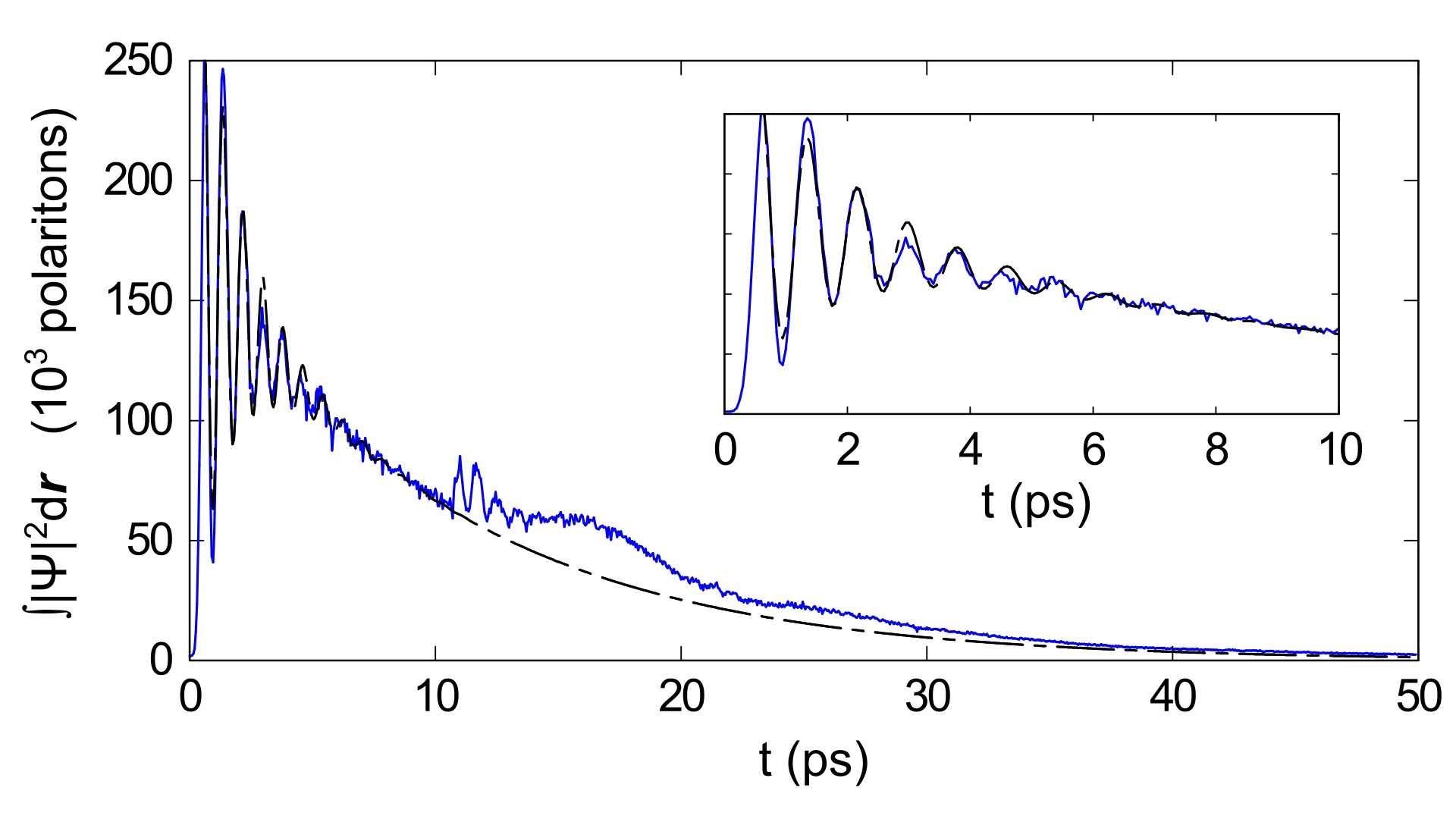}
\caption{Total intensity versus time in the femtosecond
  experiment of Fig.~1-3 of the main text. Blue line are the
  experimental data of the area-integrated emission intensity sampled
  every 50~fs. The black line is a fit based on a model of coupled and
  damped oscillators. The inset is an enlargement of the first 10~ps.}
\label{fig:backjettime}
\end{figure}

\vskip-.2cm
\begin{figure}[tbhp]
\centering\includegraphics[width=.525\linewidth]{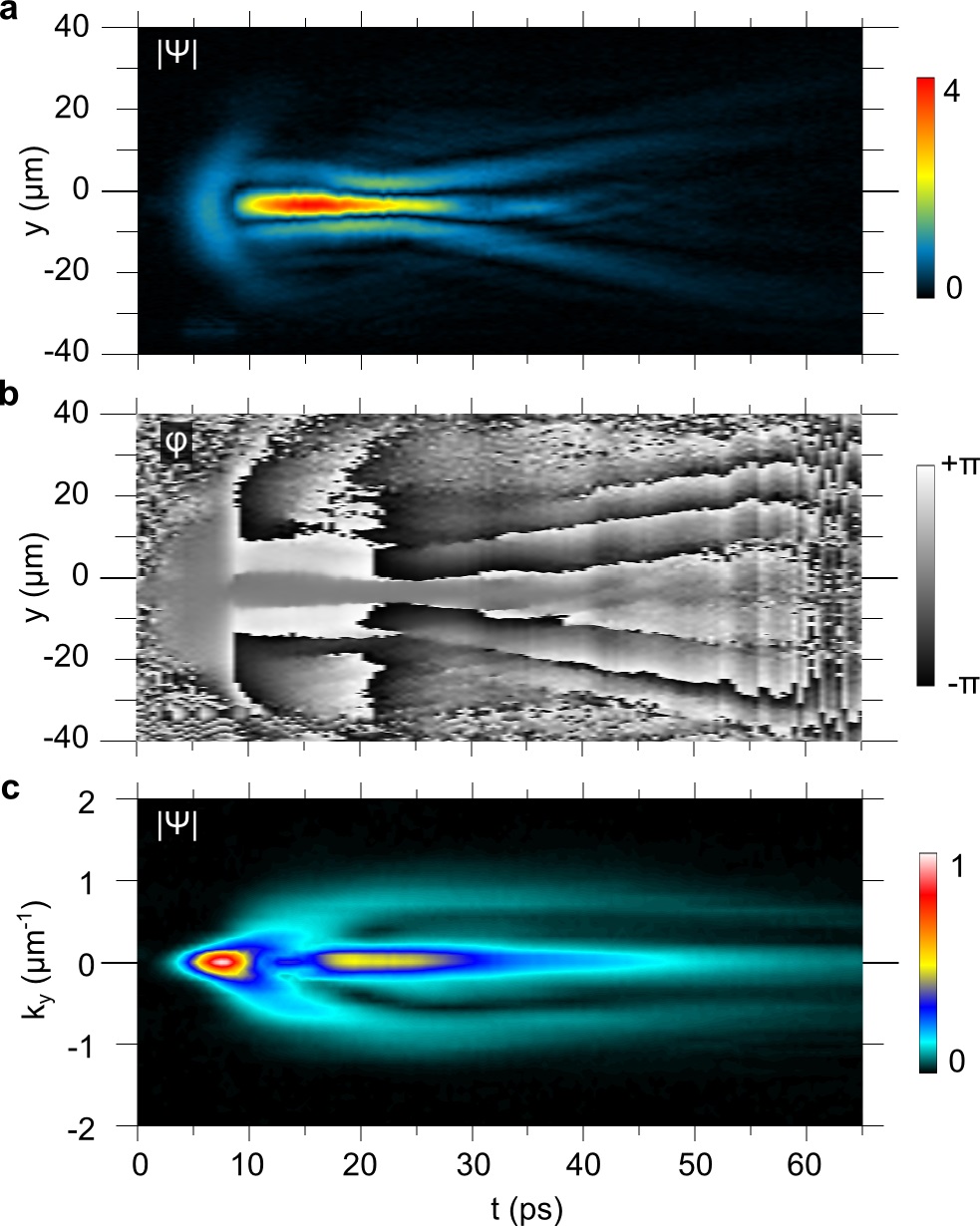}
\caption{Picosecond experiment. The excitation pulse is a
  3.5~ps width laser pulse resonant on the LPB. The three panels show
  the time-space graphs of the amplitude (a), phase (b) and $k$-space
  (c) cross cuts with a time step of 0.5~ps. The LUT colour scale in
  the case of the amplitude chart in (a) is relative to the initial
  top amplitude, and represents an enhancement factor of 4 at around
  10~ps after the pulse arrival, which corresponds to a factor of 16
  in density.}
\end{figure}

\begin{figure}[htbp]
\centering\includegraphics[width=\linewidth]{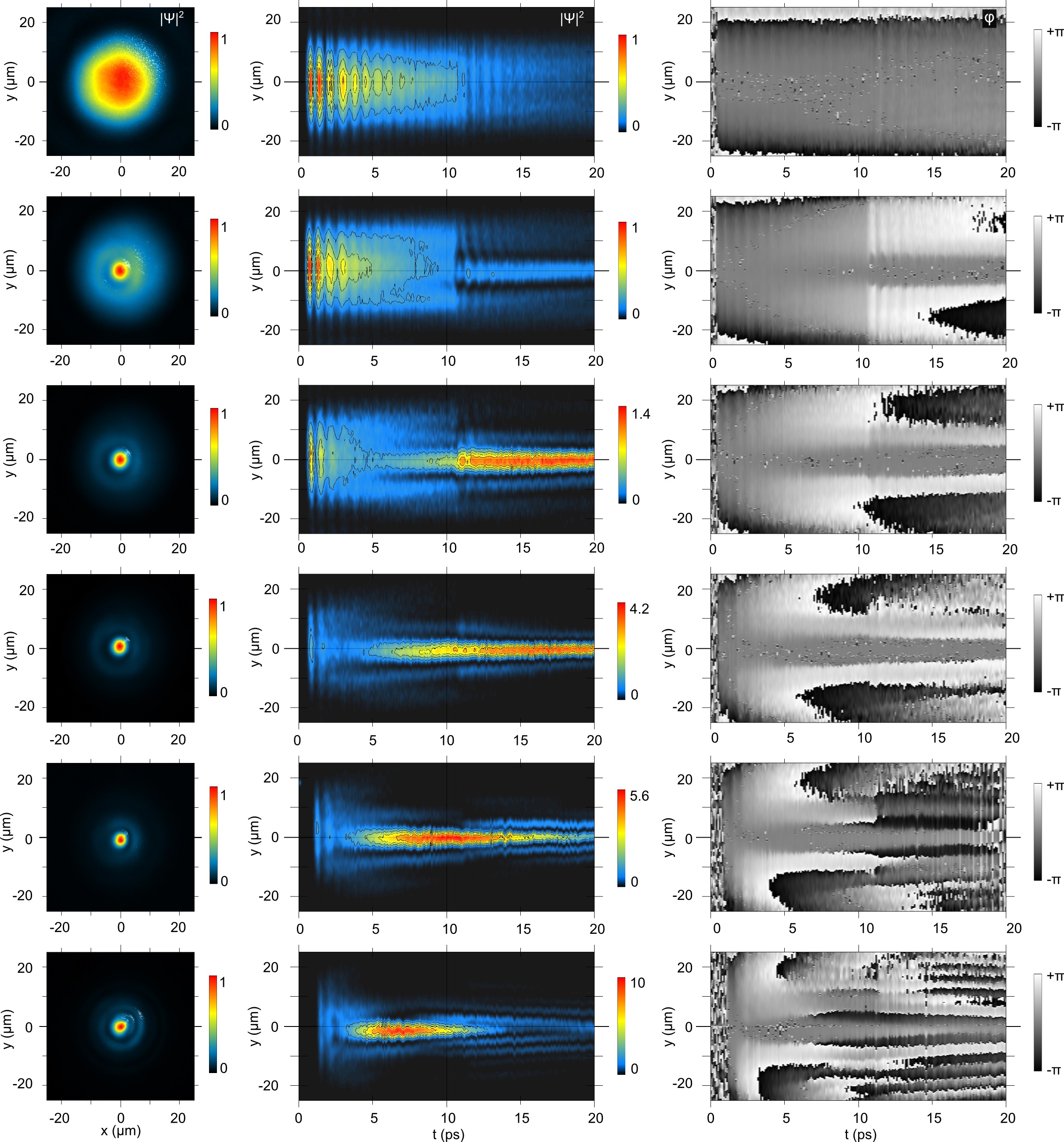}
\caption{Femtosecond experiment at several excitation powers with a
  16$\protect\mu m$ FWHM gaussian spot and linear polarization. Each
  row is relative to a different initial density. The first column
  represents the time-integrated images of the bare emission in space
  directly aquired on the camera. The second and third columns
  represent the density and phase profiles, respectively, along a
  central diameter and versus time, retrieved by means of the
  ultrafast imaging. The powers are increasing from top to bottom rows
  and are relative to initial total populations of $P_1$--$P_6$ =
  $100\times10^3$, $200\times10^3$, $350\times10^3$, $550\times10^3$,
  $1\times10^6$ and $2.1\times10^6$ polaritons. The corresponding
  initial top densities are: 345~$\mu m^{-2}$, 790~$\mu m^{-2}$,
  1200$\mu m^{-2}$, 1900$\mu m^{-2}$, 3450$\mu m^{-2}$ and 7200$\mu
  m^{-2}$, respectively. The maxima on the colour bars of the density
  charts are expressed in terms of the initial top densities and
  represent the achieved enhancement factors. From the blueshift
  associated to $P_5$, we evaluated a nonlinearity
  $g\sim0.001meV\protect\mu m^{2}$ to be renormalized to
  $g\sim0.003meV\protect\mu m^{2}$ for a single QW.}
\end{figure}

\begin{figure}[tbhp]
\centering\includegraphics[width=.4\linewidth]{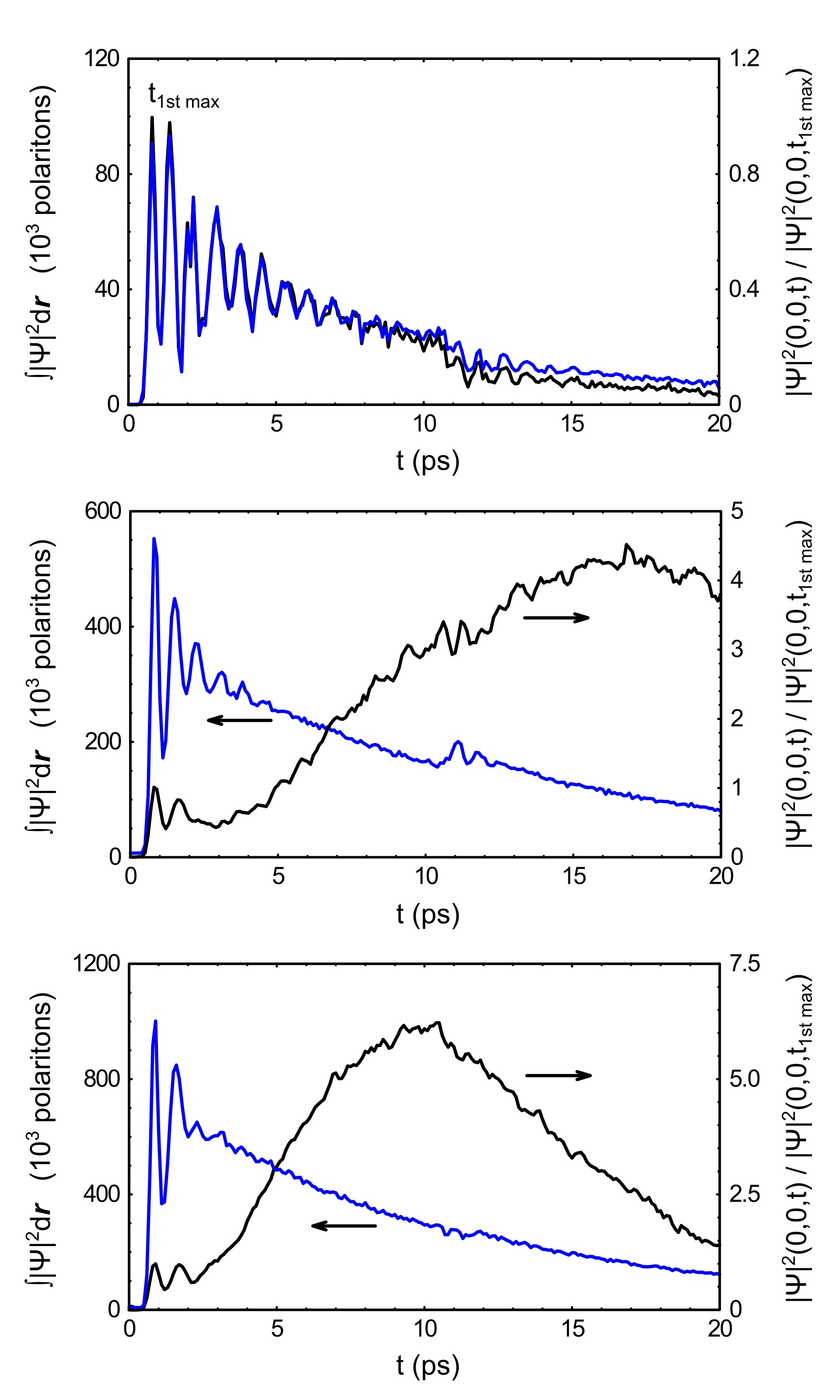}
\caption{Total and central density in the case of the femtosecond
  experiments of the previous figure S3, for the excitation
  powers~$P_1$, $P_4$ and~$P_5$ as a function of time.  The blue curve
  plotted on the left gives the total number of polaritons with time
  while the black curve plotted on the right gives the corresponding
  density in the centre as a ratio with respect to the one at the time
  of the first peak.}
\end{figure}

\vskip-2cm\begin{figure}[tbhp]
\centering\includegraphics[width=.45\linewidth]{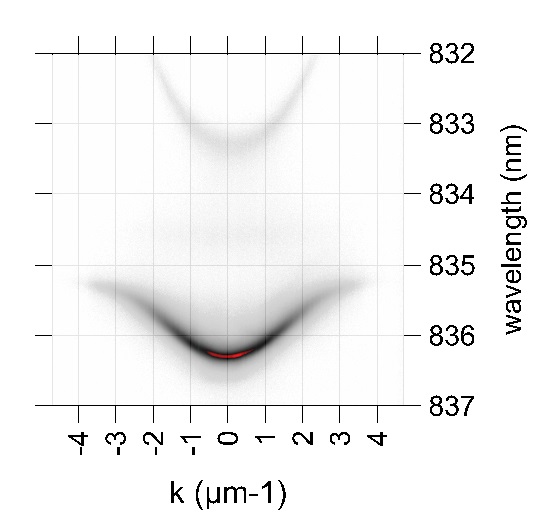}
\caption{Polariton dispersion. The bare $E$--$k$ emission of the
  microcavity polaritons is obtained after offresonant cw excitation
  at low power. The separation of the UPB and LPB branches is
  5.4~meV$/$3~nm. }
\end{figure}

\begin{figure}[htbp]
\centering\includegraphics[width=.6\linewidth]{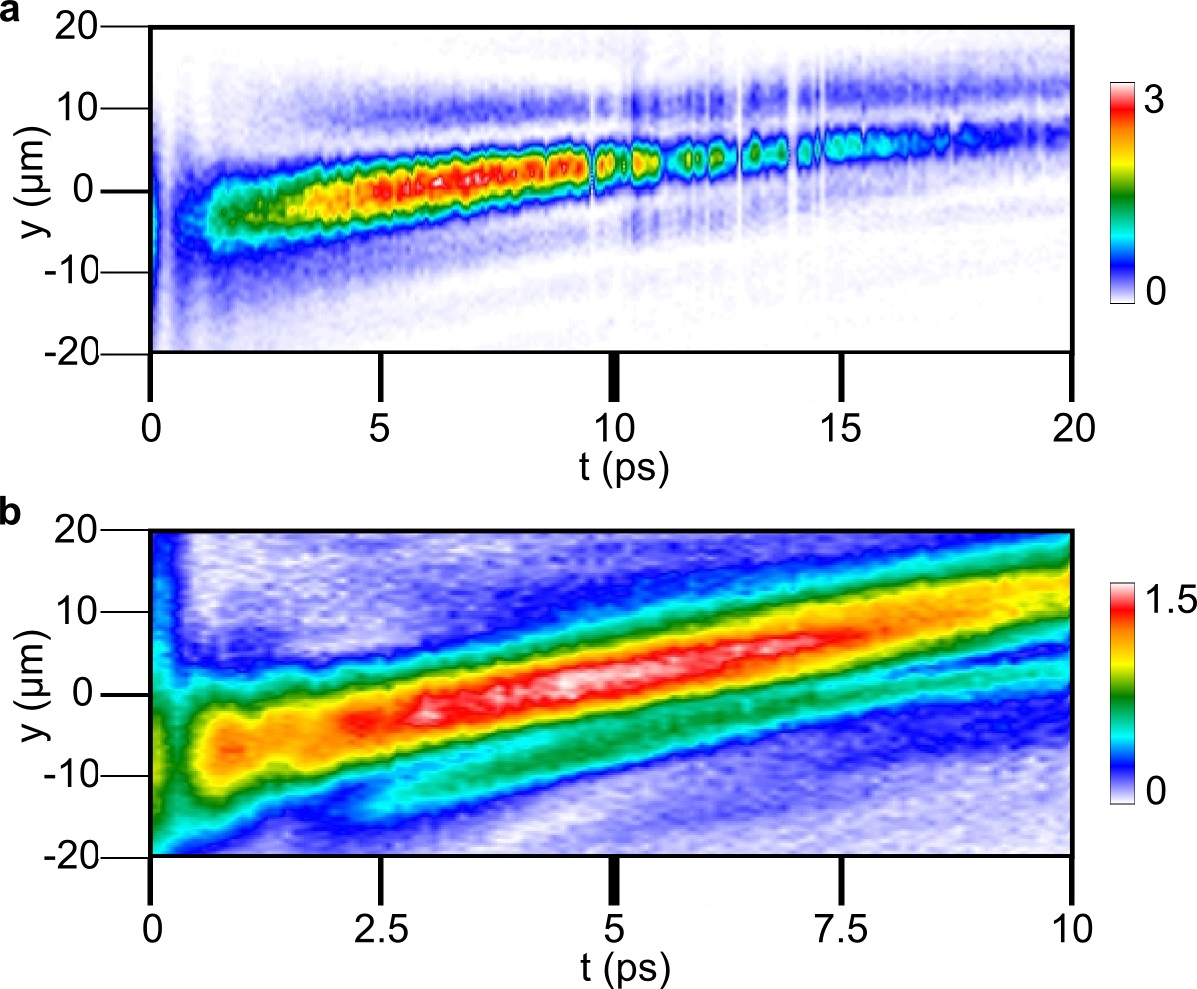}
\caption{Propagating bright peak activated under resonant fs pulse
  injection with (a) $0.7\protect\mu m^{-1}$ and (b) $1.6\protect\mu %
  m^{-1}$ in-plane wavevectors $k_y$. The cuts of the spatial
  polariton density as a function of time are taken along the
  propagation direction.}
\end{figure}

\begin{figure}[htbp]
\centering\includegraphics[width=.6\linewidth]{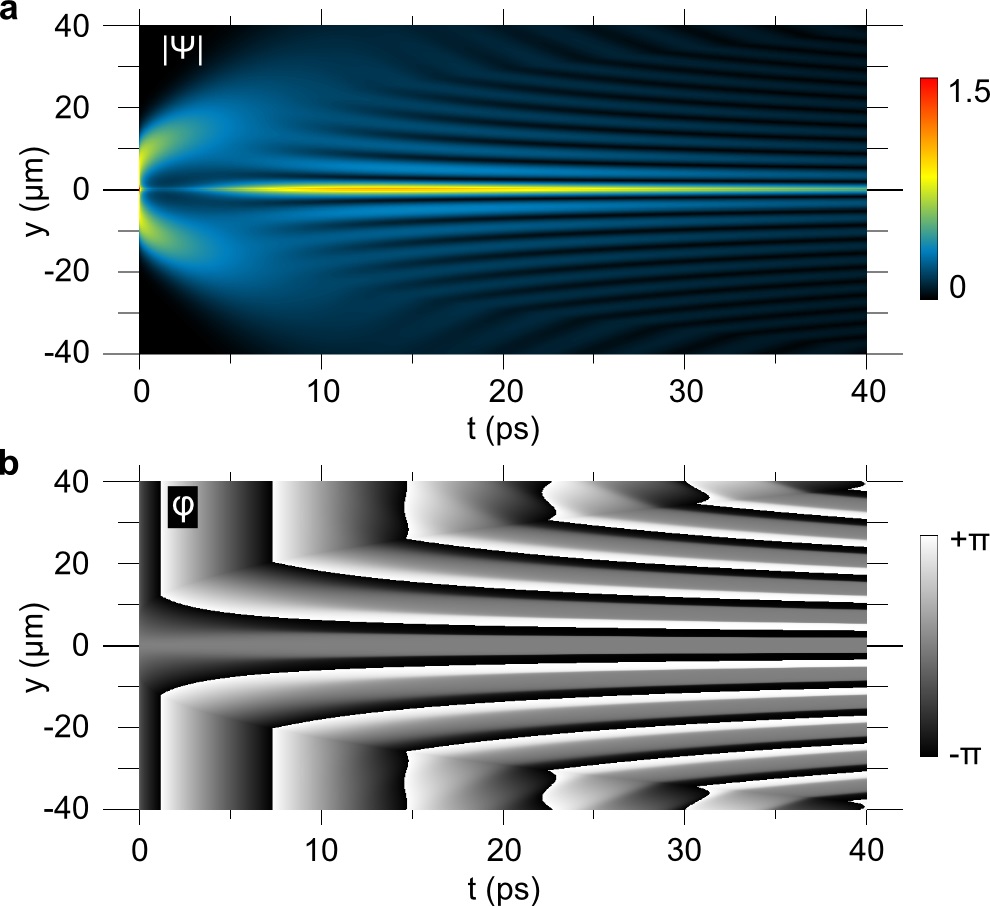}
\caption{Interferometric model from a ring source. The ring corresponds to a
  double gaussian along a diameter cross-cut.
  The ring radius and width are let to expand in time,
  which correspond to the two gaussians moving and expanding, respectively. 
  The phase is modulated by a radial in-plane
  wavevector which slightly increases in time. The part of the ring
  which reaches the centre emerges on the other side and interferes
  with the opposing wavevector fluid, causing a structure similar to
  that observed in the experiment.}
\end{figure}

\section*{Theoretical models}

The model commonly used to describe the dynamics of polaritons is
based on the mean-field approach. The time evolution of the photonic
(excitonic) wavefunctions $\phi (\mathbf{r})$ ($\chi (\mathbf{r})$) is
given by coupled Gross-Pitaevskii equations~\cite{S_Ciuti2003, S_Carusotto2004}:
\begin{equation}
  \label{eq:GPSystem}
  \begin{split}
    i\hbar \frac{\partial \phi }{\partial t}& =(E_{C}-\frac{\hbar ^{2}}{2m_{C}}%
    \nabla ^{2})\phi +\frac{\hbar \Omega _{R}}{2}\chi -\frac{i\hbar }{2\tau _{C}}%
    \phi +F_{p}(\mathbf{r},t), \\
    i\hbar \frac{\partial \chi }{\partial t}& =(E_{X}-\frac{\hbar ^{2}}{2m_{X}}%
    \nabla ^{2})\chi +\frac{\hbar \Omega _{R}}{2}\phi +g|\chi |^{2}\chi ,
  \end{split}
\end{equation}%
where $m_{C}$ ($m_{X}$) is the cavity photon (exciton) effective mass,
$g$ is the exciton-exciton interaction constant, $\Omega _{R}$ is the
Rabi frequency determining the exciton-photon coupling,
$F_{p}(\mathbf{r},t)=F_{0}e^{-\mathbf{r}^{2}/2W^{2}-t^{2}/2T_{p}^{2}-iE_{p}t}$
is the pumping field describing the pulsed excitation of the cavity
with a Gaussian spot of diameter $2W$ and pulse-duration of $2T_{p}$
and $F_{0}$ is a parameter determining the density of photo-created
carriers. While this model has been successful to describe much of the
fluid dynamics of polaritons, including ballistic
motion~\cite{S_Steger2013,S_Amo2009}, superfluidity~\cite{S_Amo2009a}, 
solitons~\cite{S_Sich2012}, etc.~\cite{S_Carusotto2013}, in
presence of repulsive interactions, $g>0$, it cannot account for
real-space localization. Many additional ingredients to these
equations can produce qualitatively the most important feature of our
experiment, the formation of a high density peak in the
center. However, these extended models, based on different physical
assumptions, have some implications, present other qualitative
features and/or require certain ranges or parameters that rule them
out as an explanation of the phenomenon. Some hypotheses that work
reasonably well demand assumptions that are difficult to justify, such
as explicit attractive interactions, $g<0$.  We found that one model
only is sufficiently consistent with all the observations and provides
a reasonably close agreement with the data to be retained as a
possible mechanism for our experiment. It is presented last in a
series of alternative descriptions, below: the loss of strong coupling
in Section~\ref{sec:viemay8085844CEST2015}, a localisation due to the
exciton reservoir or dark excitons in
Section~\ref{sec:viemay8121915CEST2015}, polaritons with attractive
interactions in Section~\ref{sec:viemay8123718CEST2015}---all of these
models failing in some fatal way to account for the experiment---and,
ultimately, the collective polaron effect, in
Section~\ref{sec:viemay8130427CEST2015}, which, on the contrary,
reproduces adequately the findings, based on likely assumptions and
parameters corresponding to our sample.

\section{Loss of strong-coupling}
\label{sec:viemay8085844CEST2015}

\begin{figure}[tbp]
\qquad\includegraphics[width=7cm]{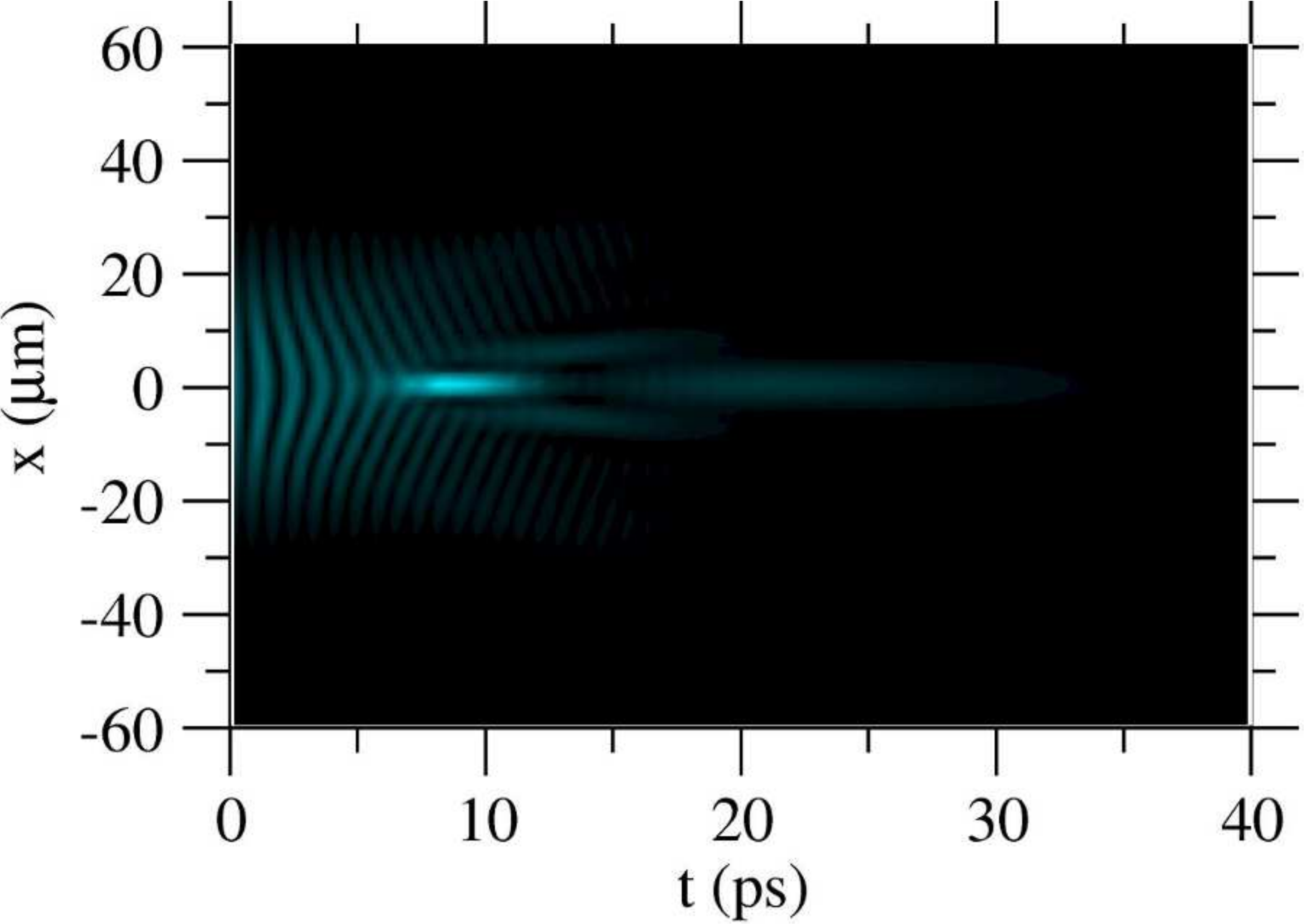} \qquad %
\includegraphics[width=7cm]{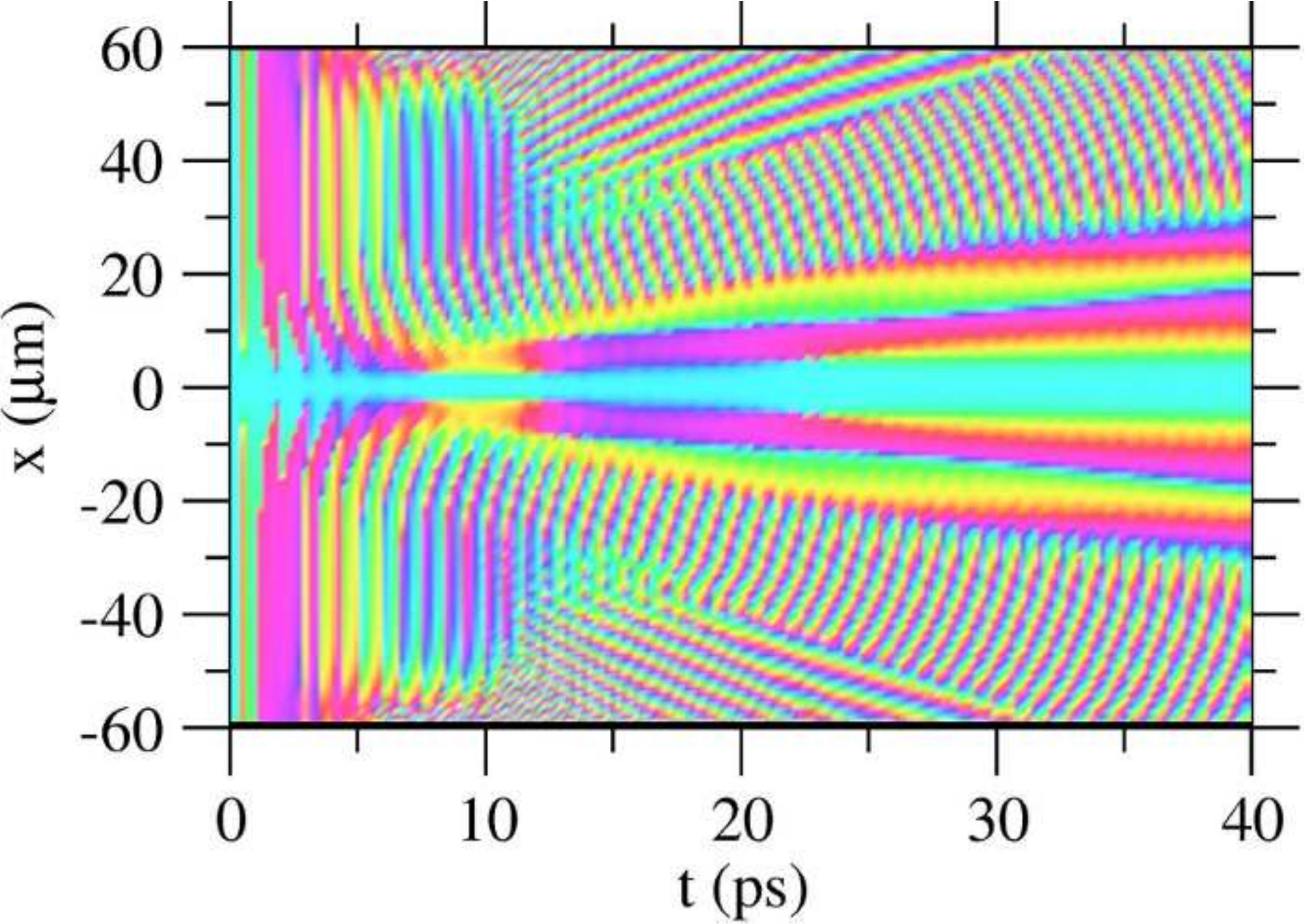} \newline
\includegraphics[width=7cm]{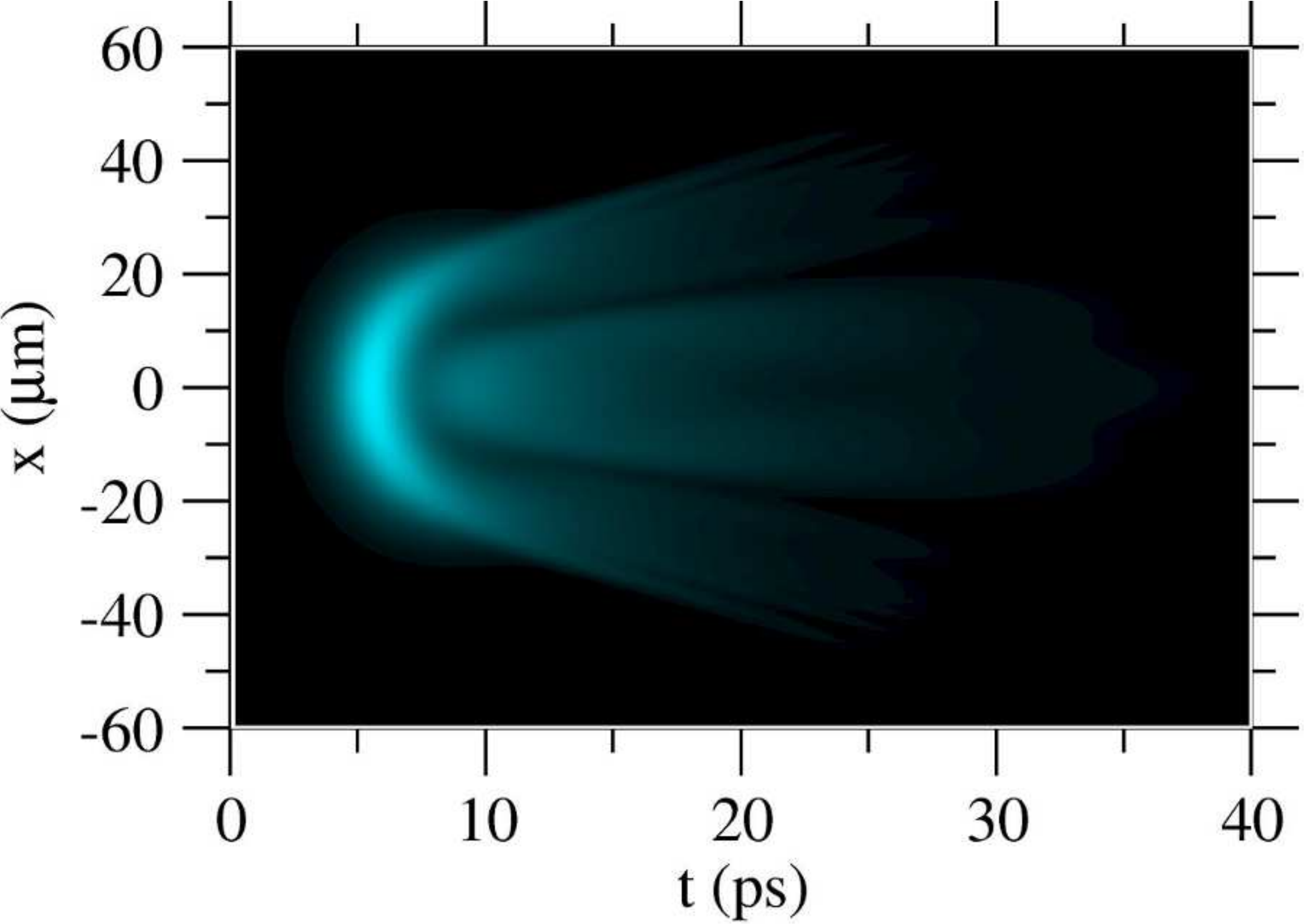} \qquad %
\includegraphics[width=7cm]{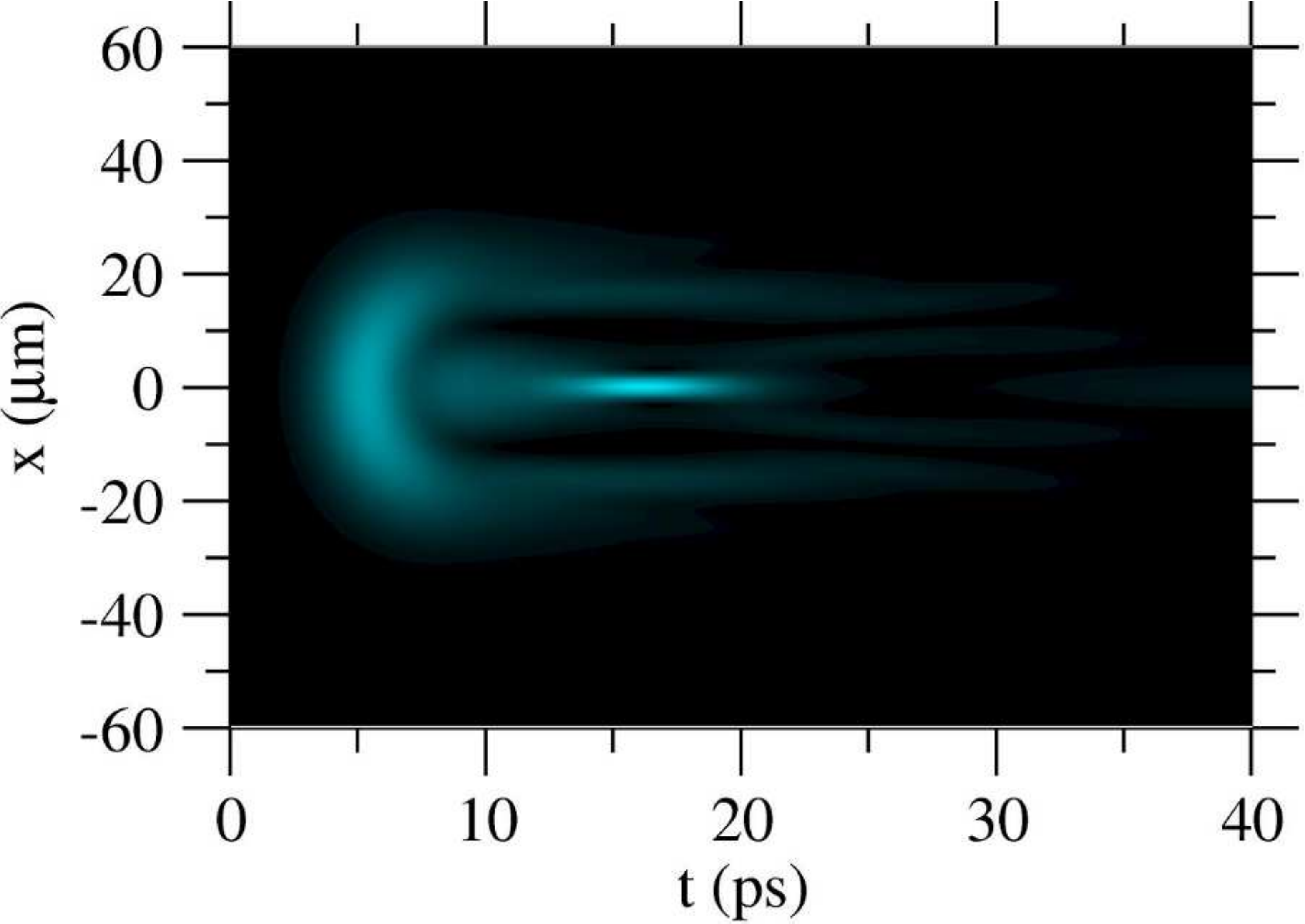}
\caption{Evolution of the photon field intensity in the model with
  density dependent reduction of the Rabi coupling. Top frames show
  the photon field density and phase, respectively, in the case of
  femtosecond pulse excitation. Bottom frames show picosecond pulse
  excitation at the position of the UP branch (left), featuring a
  real-space localization, and at the position of the LP branch
  (right), failing to produce the localization. Parameters are
  $\Omega_R=2\protect\pi/(0.8 \mathrm{ps}) - (3/2\hbar) g
  |\protect\chi|^2$%
  , $\protect\delta=-0.5$ meV, $m_C=3\times 10^{-5} m_e$,
  $\protect\tau_p=4$ ps, $\protect\tau_x=300$ ps,
  $\protect\gamma,\protect\eta=0$, $g=2\times 10^{-2}$ meV
  $\protect\mu$m$^2$, $\tilde{g}=g$, $\hbar R=2.6\times 10^{-3}$ meV
  $\protect\mu$m$^2$, $W=12.5 \protect\mu$m, $T_{pulse}=50$ fs,
  $\Delta \protect\omega=0$, $n_R= 0$, $W=12.5 \protect\mu$m,
  $T_{pulse}=50$ fs (fs-pulse case) or $T_{pulse}=2$ ps (ps-pulse
  case). }
\label{fig:Mikhail}
\end{figure}

The measured emission of total emitted photons from a single
experiment suggests that the exciton density is close to the
saturation value of $n_\mathrm{sat}\approx 10^{3}\mu m^{-2}$. At such
a high density, the reduction of the Rabi splitting should occur due to
phase space
filling~\cite{S_Schmitt-Rink1985,S_Rochat2000,S_Ciuti2003,S_Kwong2001,S_Luk2013} and lead
to a modification of the polariton dispersion. At the first order of
perturbation in the exciton density, the saturation is
momentum-independent, or local in space (precisely, the range of the
interaction is of the order of the exciton
radius)~\cite{S_Rochat2000,S_Kwong2001,S_Luk2013}. The linear regime Rabi
coupling~$\Omega_{R0}$ is then renormalized to:
\begin{equation}
  \label{eq:viemay8103312CEST2015}
  \Omega_R = \Omega_{R0} - a \frac{g |\chi|^2}{\hbar}
\end{equation}
where~$a$ is the coefficient dependent on the quantum well geometry,
$g$ the polariton-polariton interaction strength and~$\chi$ the
exciton wavefunction (as already defined before).  This leads to
effective attractive interactions but only for the UP branch, while
the interactions remain repulsive for the LP branch~\cite{S_Ciuti2003}.
This is because the UP branch is redshifted whilst the Rabi coupling
decreases, while the LP branch is blueshifted, just as it would be
from repulsive interactions. A numerical simulation for this mechanism
is shown in Fig.~\ref{fig:Mikhail}. This model correctly reproduces
the experimental Rabi bending and allows to describe a central density
peak with both a femtosecond pulse and a picosecond pulse tuned to the
UP branch. Critically, however, the ps pulse excitation tuned to the
LP branch leads to strong defocusing in this model, in contradiction
with the experimental data.

\begin{figure}[tbp]
\includegraphics[width=8cm]{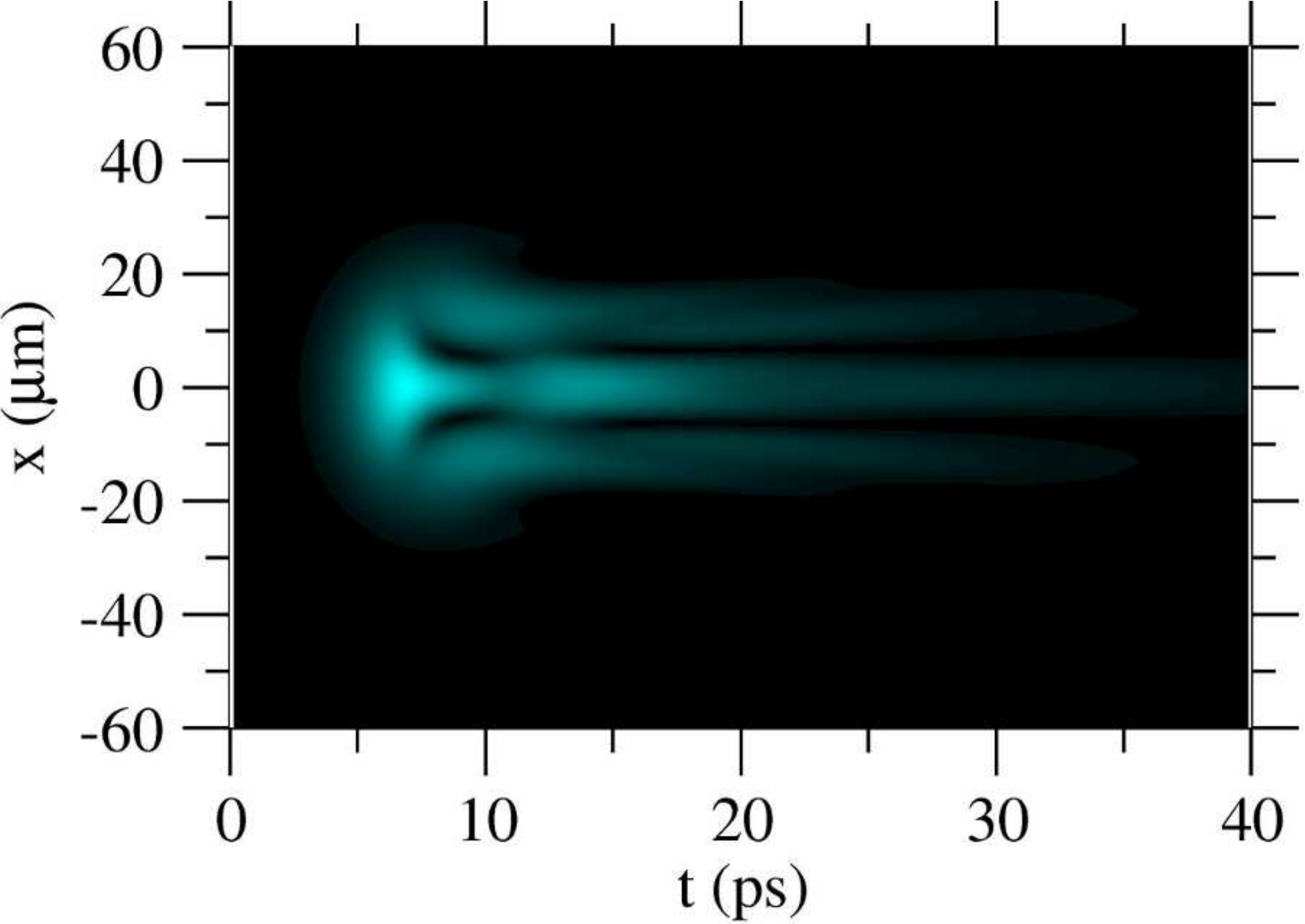}
\caption{Evolution of the photon field intensity in the model with
k-dependent nonlocal reduction of the Rabi coupling. Parameters are the same
as in Fig.~\protect\ref{fig:Mikhail} (for ps pulse excitation at the LP
branch) with $a=4$ and $k_c=0.2 \protect\mu$m$^{-1}$. }
\label{fig:negative}
\end{figure}

To overcome the defocusing in the case of LP branch excitation, we can
consider a momentum-dependent loss of the Rabi coupling.  Assuming a
stronger reduction of $\Omega_R$ in the vicinity of $k=0$, where lies
the condensate whose density is responsible for the phase-space
filling, allows to create a negative effective mass for the lower
polariton.  Indeed, the LP polariton branch blueshifts more at $k=0$,
creating a negative curvature of the dispersion around the ground
state, allowing for self-focusing even with repulsive
interactions. Technically, this can be achieved by filtering
Eq.~(\ref{eq:viemay8103312CEST2015}) in $k$ space. In the simplest
formulation, this is done by transforming the excitonic wavefunction
$\phi = |\chi|^2 \psi$ to $k$-space, applying a Gaussian filter
$\phi_1(k) = \phi(k) \exp(- k^2 / 2 k_c^2)$, transforming it back, and
subtracting from the $\Omega_{R0}$ term after multiplication by $a
g/\hbar$. This nonlocal and nonlinear reduction of the Rabi coupling
is equivalent to introducing interactions that are nonlocal in real
space. The range of the nonlocal effective interaction is given by
$k_c^{-1}$. Figure~\ref{fig:negative} shows an example of
self-focusing in this model in the case of picosecond LP branch
excitation. In the case of femtosecond excitation, the results are
similar to the ones obtained with the previous model.

While this model provides a good agreement with experiments performed
at different settings, its justification is not straightforward. While
the $k$-dependent reduction of Rabi coupling can be predicted by the
simple model of phase space filling or exchange
effects~\cite{S_Schmitt-Rink1985}, it occurs at $k$ vectors comparable to
the inverse exciton Bohr radius $a_B^{-1}$. The dependence of the
$\Omega_R$ on momentum for $k$ comparable to $\mu$m$^{-1}$ requires an
existence of correlations over distances much larger than the exciton
radius, for which we see no rationale in the mean-field framework.  An
explanation along these lines is thus thwarted in a straightforward
model and would require a drastic reconsideration of the physics of
polaritons based on the mean-field approximation and perturbation
theory in the exciton basis. In the close vicinity of the Mott
transition, the description in terms of the dilute exciton gas may
become questionable, and collective effects of strongly-correlated
quantum gases could be required instead. Indeed, BCS-related physics
is expected to occur close to the phase space filling
threshold~\cite{S_Keeling2005}. Such an interpretation cannot be
excluded and would open a new page of the field, but it is, however,
much beyond the scope of this text.

\section{Localisation by the reservoir}
\label{sec:viemay8121915CEST2015}

The exciton reservoir is another possible candidate to explain the
redistribution of the polariton condensate.  A conceivable
scenario involves the creation of a ring of reservoir excitons around
the center of the pumping spot, which subsequently creates a backflow
of polariton waves, interfering constructively in the center and
giving rise to the strong intensity peak. Such a model consists of the
photon, exciton, and reservoir fields, similar to
Ref.~\cite{S_Wouters2007}, but treating coherent photons
and excitons with separate fields:
\begin{align}  \label{reservoir_ring}
i \hbar\frac{\partial \phi}{\partial t} &=-(1-iA)\frac{\hbar^2 }{2 m_C}
\nabla^2 \phi +\frac{\hbar \Omega_R}{2}\chi -\frac{i\hbar}{2\tau_p} \phi + F(%
\mathbf{r},t)\,,  \notag \\
i \hbar\frac{\partial \chi}{\partial t} &= \frac{\hbar \Omega_R}{2}\phi - i
\delta \chi + g |\chi|^2 \chi + \tilde{g} n_{\mathrm{R}} \chi +i\frac{\hbar}{%
2}\left(R n_{\mathrm{R}} - \frac{1}{\tau_x} \right) \chi\,, \\
\frac{\partial n_{\mathrm{R}}}{\partial t} &= -\left(\frac{1}{\tau_x} + R
|\chi|^2\right) n_{\mathrm{R}}\,,  \notag
\end{align}
where $A$ is the energy relaxation constant, $F(%
\mathbf{r},t) = F_0 \mathrm{e}^{-\mathbf{r}^2/2W^2 -t^2/2T_{pulse}^2 -
  i\Delta \omega t}$ is the pumping field and with an initial
reservoir density $n_R(\mathbf{r},t=0)=
n_0\mathrm{e}^{-\mathbf{r}^2/2W^2}$. In this simple model, the
backscattering from the condensate to the reservoir is not taken into
account, but the initial reservoir density is assumed to be excited by
the incoming pulse.

\begin{figure}[tbp]
\includegraphics[width=8cm]{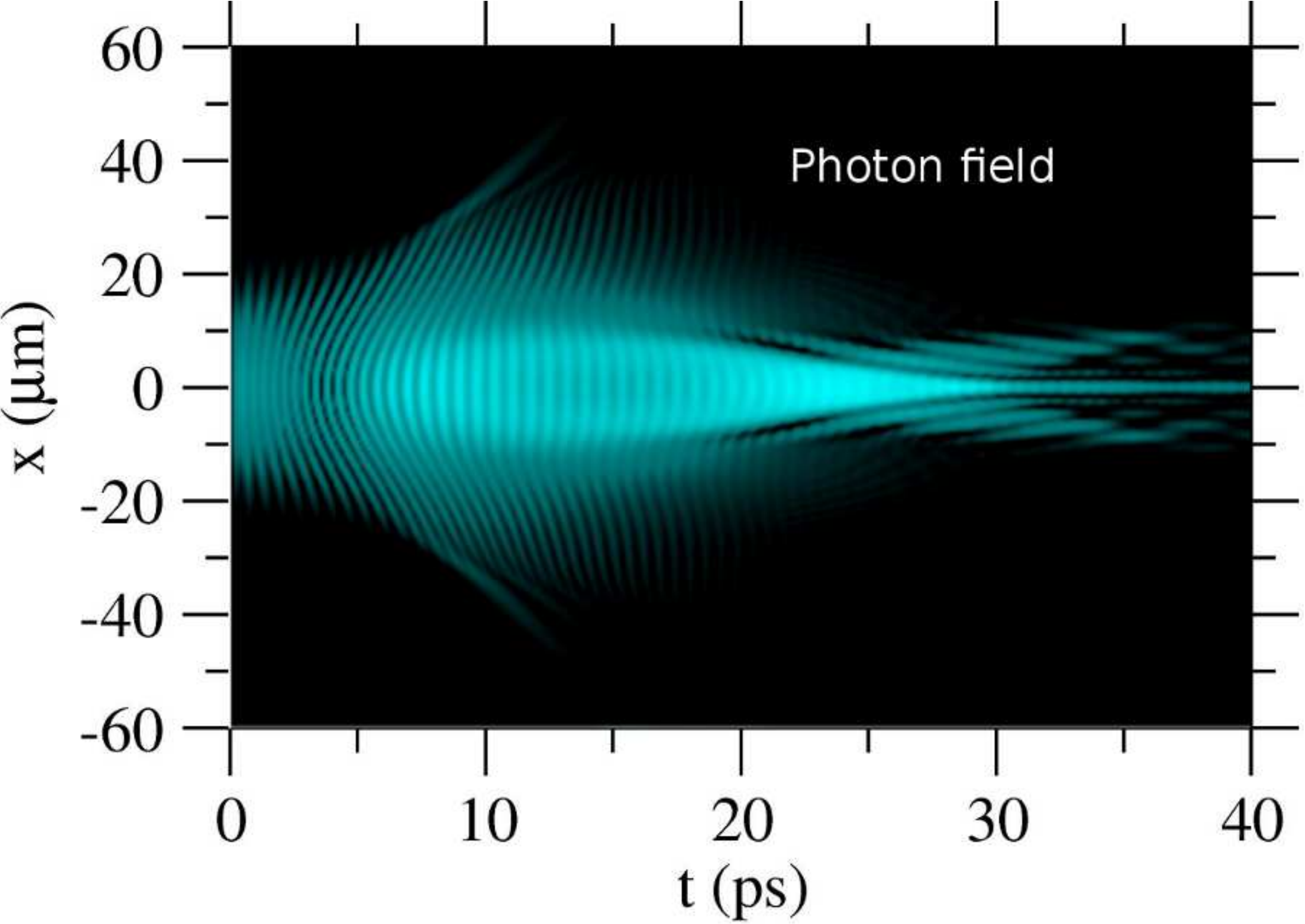} \qquad %
\includegraphics[width=8cm]{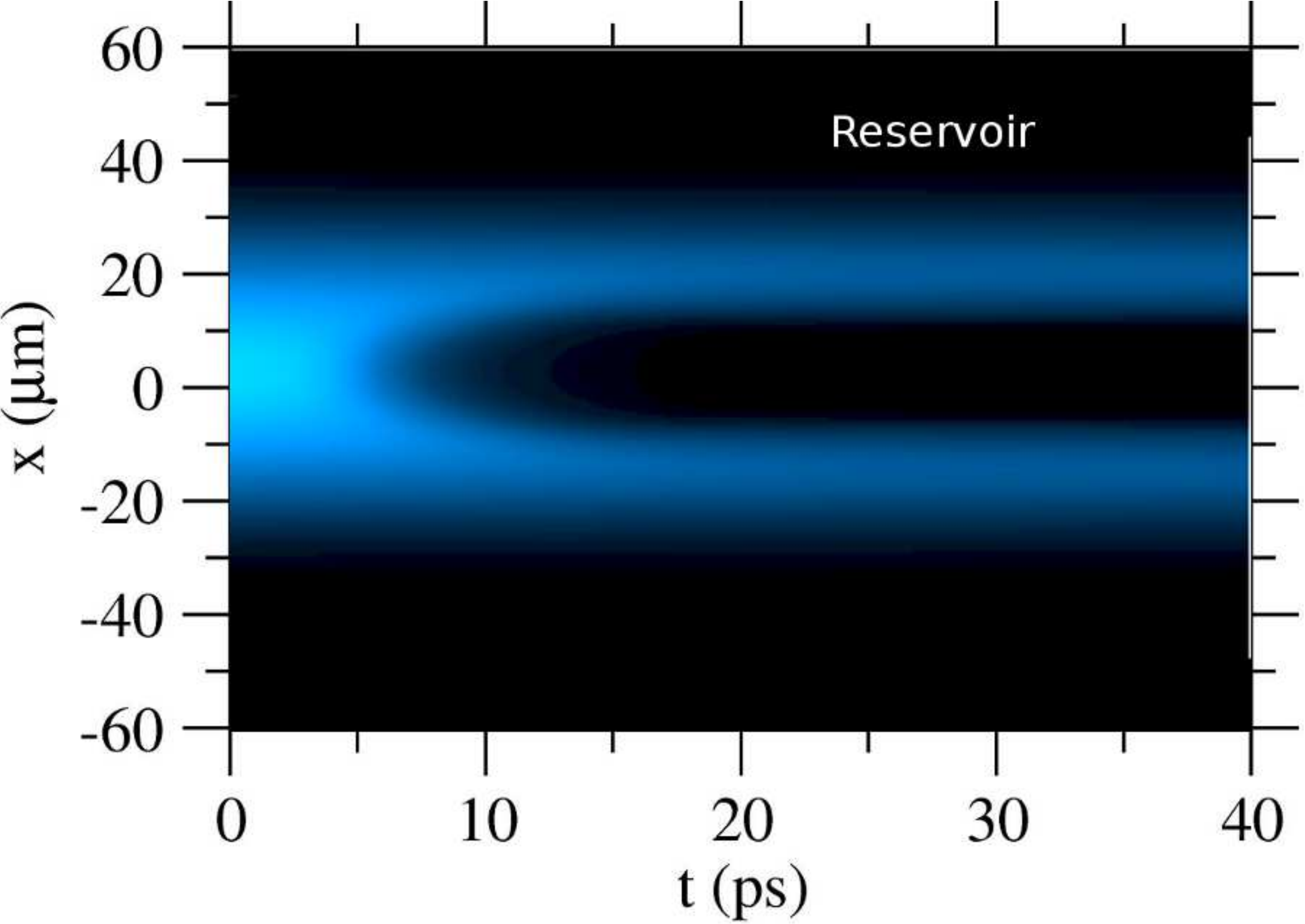}
\caption{Evolution of the photon field intensity and reservoir density in
the reservoir-ring model. Parameters are $A=0$, $\protect\delta=0$, $%
m_C=5\times 10^{-5} m_e$, $\hbar\Omega_R=6$ meV, $\protect\tau_p=2.5$ ps, $%
\protect\tau_x=300$ ps, $g=12\times 10^{-3}$ meV $\protect\mu$m$^2$, $\tilde{%
g}=2g$, $\hbar R=2.6\times 10^{-3}$ meV $\protect\mu$m$^2$, $W=12.5 \protect%
\mu$m, $T_{pulse}=50$ fs, $\Delta \protect\omega=0$. }
\label{fig:reservoir}
\end{figure}

A numerical simulation of this model is shown in
Fig.~\ref{fig:reservoir}. The depletion in the center of the reservoir
is created due to the stimulated scattering to the condensate, which
is faster in the center, where the condensate density is larger. The
bright density peak is then created because the polariton waves
emitted form the remaining ring-shaped reservoir ``focus'' in the
center. While the main feature of the experiment is thus reproduced,
there are several problems with quantitative features: (a) the central
peak appears much later than is observed in the experiment and (b) the
bending of the Rabi oscillations is initially in the incorrect
direction (backwards in the center).  There are even more serious
conceptual difficulties: (c) in such a model, the peak cannot move
``ballistically'', as is observed in the experiment if the excitation
pulse is injected at an angle, since the reservoir would stay
essentially in the position where it is created.  (d) The peak
persists when resonantly exciting the lower branch, which is expected
to produce little or no reservoir excitons. (e) Finally, the effect
does not show any strong polarization dependence.
In the regime of quasiresonant pulsed excitation we expect the reservoir to be polarised as the
typical exciton spin-relaxation time in our system is supposed to be long 
compared to the characteristic time scale of the Rabi-oscillations and collapse dynamics

Variations on this theme are possible. For instance, one can assume
the intermission of dark excitons. Since the coupling is incoherent
and assuming the exciton mass as infinite, which allows to treat dark
excitons as a density field, their effect can be modeled by the
following equations:
\begin{align}
i\hbar \frac{\partial \psi }{\partial t}& =\left( -\frac{\hbar ^{2}}{2m_{LP}}%
\nabla ^{2}-\frac{i\hbar }{2\tau _{p}}-i\gamma |\psi |^{2}+i\gamma _{d}n_{%
\mathrm{dark}}^{2}+g\left( |\psi |^{2}+n_{\mathrm{dark}}\right) \right) \psi
+F(\mathbf{r},t),  \label{Alexey2} \\
\frac{\partial n_{\mathrm{dark}}}{\partial t}& =\gamma |\psi |^{4}-\gamma
_{d}n_{\mathrm{dark}}^{2}|\psi |^{2}.  \notag
\end{align}

\begin{figure}[tbp]
\includegraphics[width=8cm]{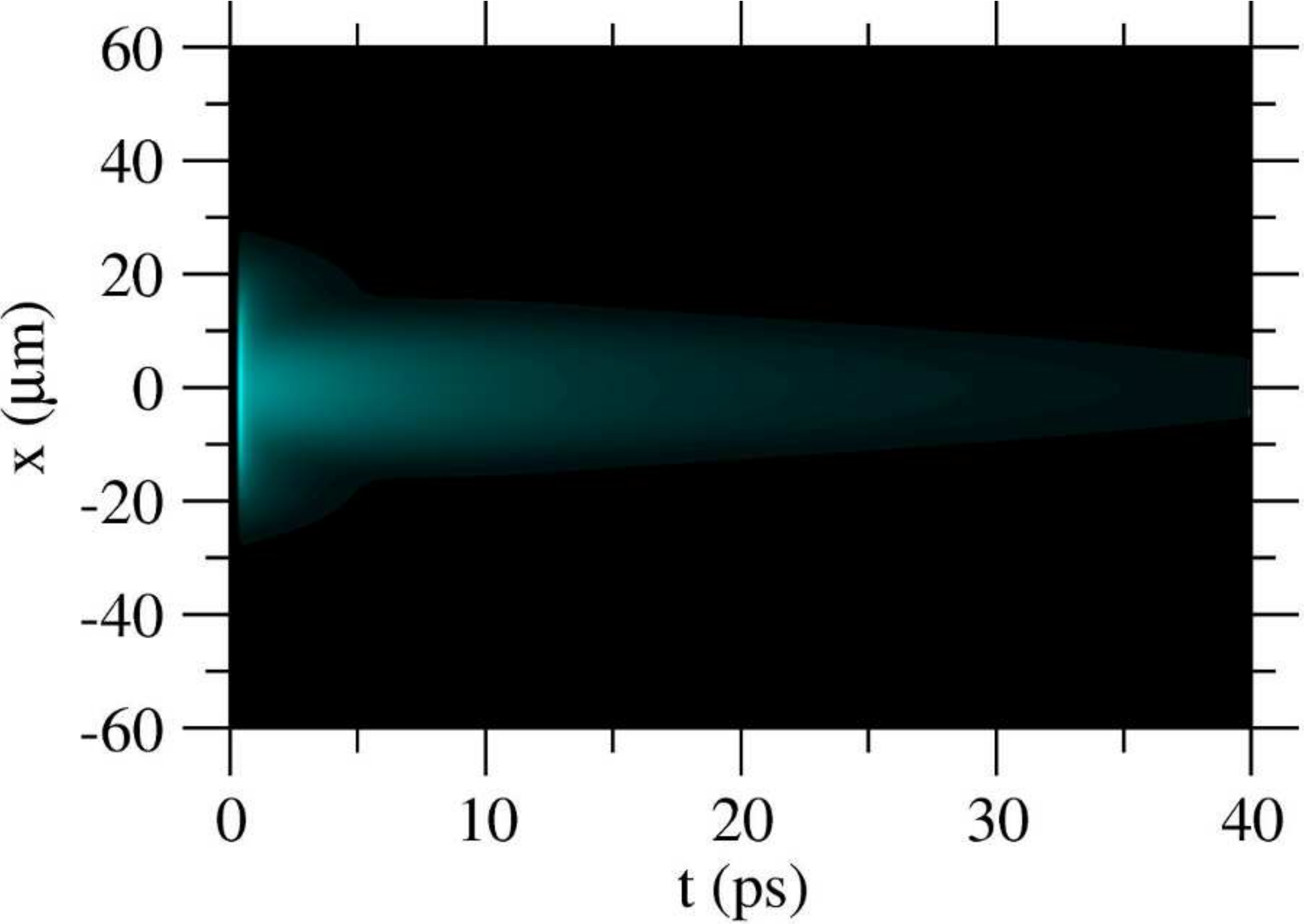}
\caption{Evolution of the photon field intensity in the model with dark
excitons. Parameters are the same as in Fig.~\protect\ref{fig:reservoir},
except $m_{LP}=3\times 10^{-5} m_e$, $\protect\tau_p=4$ ps, $\protect\gamma%
_x=0.4$ $\protect\mu$m$^2$ ps$^{-1}$, $\protect\gamma_d=0.02$ $\protect\mu$m$%
^4$ ps$^{-1}$, $g=2\times 10^{-2}$ meV $\protect\mu$m$^2$, $n_0= 0$.}
\label{fig:Alexey}
\end{figure}

A numerical simulation of this model is shown in
Fig.~\ref{fig:Alexey}.  The scattering to polaritons creates a hole in
the Gaussian distribution of excitons. The flow towards the hole can
create a localization, however, the effect is not as clear as in the
experiment: the localization is rather broad and has a low intensity.

\section{Attractive interactions between excitons}
\label{sec:viemay8123718CEST2015}

\begin{figure}[tbp]
\includegraphics[width=8cm]{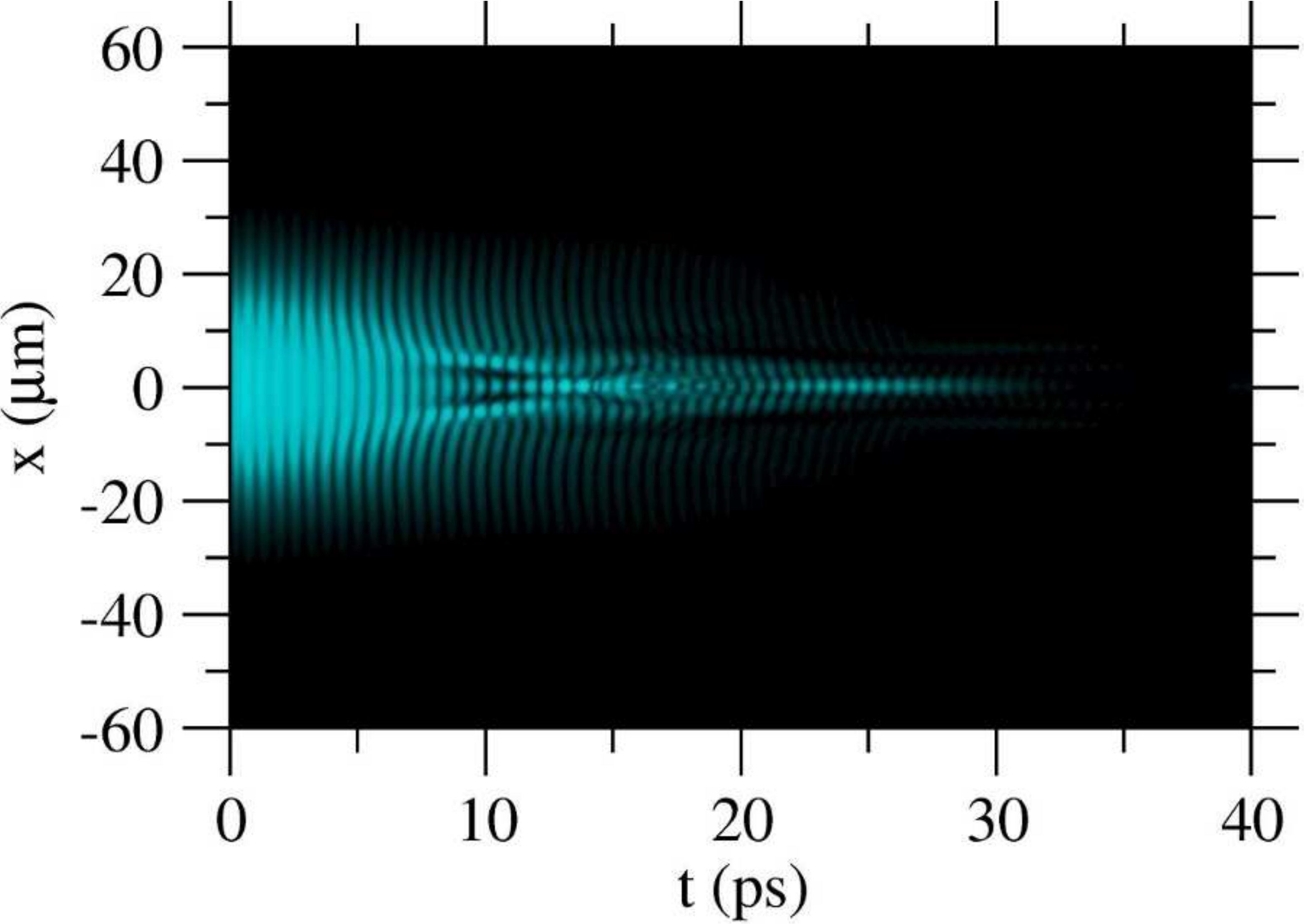}
\caption{Evolution of the photon field intensity in the model with
attractive polariton-polariton interactions. Parameters are the same as in
Fig.~\protect\ref{fig:reservoir}, except $A=0.1$, $\protect\delta=3$ meV, $%
g=-12\times 10^{-3}$ meV $\protect\mu$m$^2$, $\tilde{g}=2|g|$, $\hbar
R=4.6\times 10{-3}$ meV $\protect\mu$m$^2$. }
\label{fig:attractive}
\end{figure}

The interaction constant in the Gross-Pitaevskii equations, and
variants, that describe the polariton dynamics is positive, as indeed
same-spin excitons are largely believed to be repulsive, but several
effects could potentially give rise to attractive interactions, as we
discuss shortly. Since we actually observe a real-space collapse, it
is instructive to simply assume attractive interactions between
polaritons.  In this case self-focusing is indeed possible for a
positive effective mass. The model is identical to the one above,
Eqs.~(\ref{eq:GPSystem}), but with $g<0$.  We do not include the
reservoir excitons that we have seen can be ruled out on several
grounds. A result of numerical simulation is shown in
Fig.~\ref{fig:attractive}.  As expected, the high intensity peak
appears due to self-focusing.

While some features of the experiment are well reproduced in this
model, the existence of attractive polariton interactions is highly
questionable. Potentially the strongest effect in this sense is the
indirect exchange through the biexciton state, but this occurs only
for opposite spins, while we did not observe any strong spin
dependence of the focusing effect in the experiment. The other
possible sources of attractive interaction, such as the indirect
exchange~\cite{S_Ciuti1998,S_Vishnevsky2014,S_Vladimirova2010},
or Van der Waals interactions appear to be negligible for the
low-momentum excitons created by resonant pumping~\cite{S_Vladimirova2010}.
It was also recently observed that the sign of the interactions may
change to negative at large polariton
momenta~\cite{S_Vishnevsky2014}, and other effects could
similarly give rise to polariton
attraction~\cite{S_Vishnevsky2014,S_Vladimirova2010}, but the proposed
configurations are difficult to fit in with our experiment. The change
of sign of the polariton interactions with no further justification is
thus not tenable and would be in conflict with a large literature that
successfully describes various experimental regimes on a physically
well motivated assumption of repulsive interactions.

\section{Collective polaron effect}
\label{sec:viemay8130427CEST2015}

We now conclude this series of possible models to reproduce the
real-space collapse of the polariton condensate with the one model
that we found could stand as an explanation for the observed effect.

When the exciton density is large enough, as in the case of pulsed
excitation, the macroscopic occupation of the ground state with
excitons may lead to enhancement of the exciton-phonon interaction
that produces a cooling of the excitonic reservoir against its
blueshift due to their attraction~\cite{S_Ivanov1999,S_Porras2002}. The
corresponding released energy is transfered to the crystal lattice via
acoustic phonons. Another plausible mechanism of heating is through
the polariton-polariton Auger process followed by emission of the
cascade of acoustic phonons as discussed by Klembt~\textit{et
  al.}~\cite{S_Klembt2015}. In any case, a local heating may occur which
leads to a band-gap narrowing, a variation of the exciton-photon
detuning and a redshift of the lower polariton branch. The latter
effect may lead to self-focusing of the polariton condensate. Such an
effect can be referred to as a ``collective polaron
effect''. Formally, it can be described by the modulation of the
exciton energy through a retarded function:
\begin{equation}
  \label{eq:EX}
  E_{X}(t)=-\frac{\beta }{\tau _{X}}\int_{-\infty }^{t}\left\vert \chi \left( 
      \mathbf{r},t^{\prime }\right) \right\vert ^{4}e^{-\frac{t-t^{\prime }}{\tau
      _{X}}}dt^{\prime },  
\end{equation}
where $\beta $ characterises the magnitude of band-gap renormalization
and $\tau_{X}$ is the characteristic heat relaxation time in the
crystal lattice.  Both $\beta$ and $\tau_{X}$ are fitting
parameters. The fourth power of the exciton wavefunction
$\chi\left(\mathbf{r},t^{\prime }\right)$ under the integral accounts
for the quadratic dependence of the polariton Auger process on the
concentration of polaritons.  The dynamics of the system is governed
by the initial pumping strength. The Rabi oscillations generally have
an anharmonic behavior due to the changes of the bare exciton energy
and the interplay between the blueshift $g|\chi|^{2}$ and the redshift
$E_{X}(t)$. The effective frequency of these oscillations is then
determined by $\Omega _{\mathrm{eff}}=\sqrt{\Omega _{R}^{2}+(\Delta
  -g|\chi |^{2})^{2}}$~\cite{S_Voronova2015}. In addition to the
change in detuning, at larger population densities close to the Mott
transition, the coupling $\Omega_{R}$ itself would be reduced,
resulting from the change of the exciton oscillator strength due to
the exciton phase space filling~\cite{S_Schmitt-Rink1985}. 
\begin{figure}[tbp]
\includegraphics[width=.6\columnwidth]{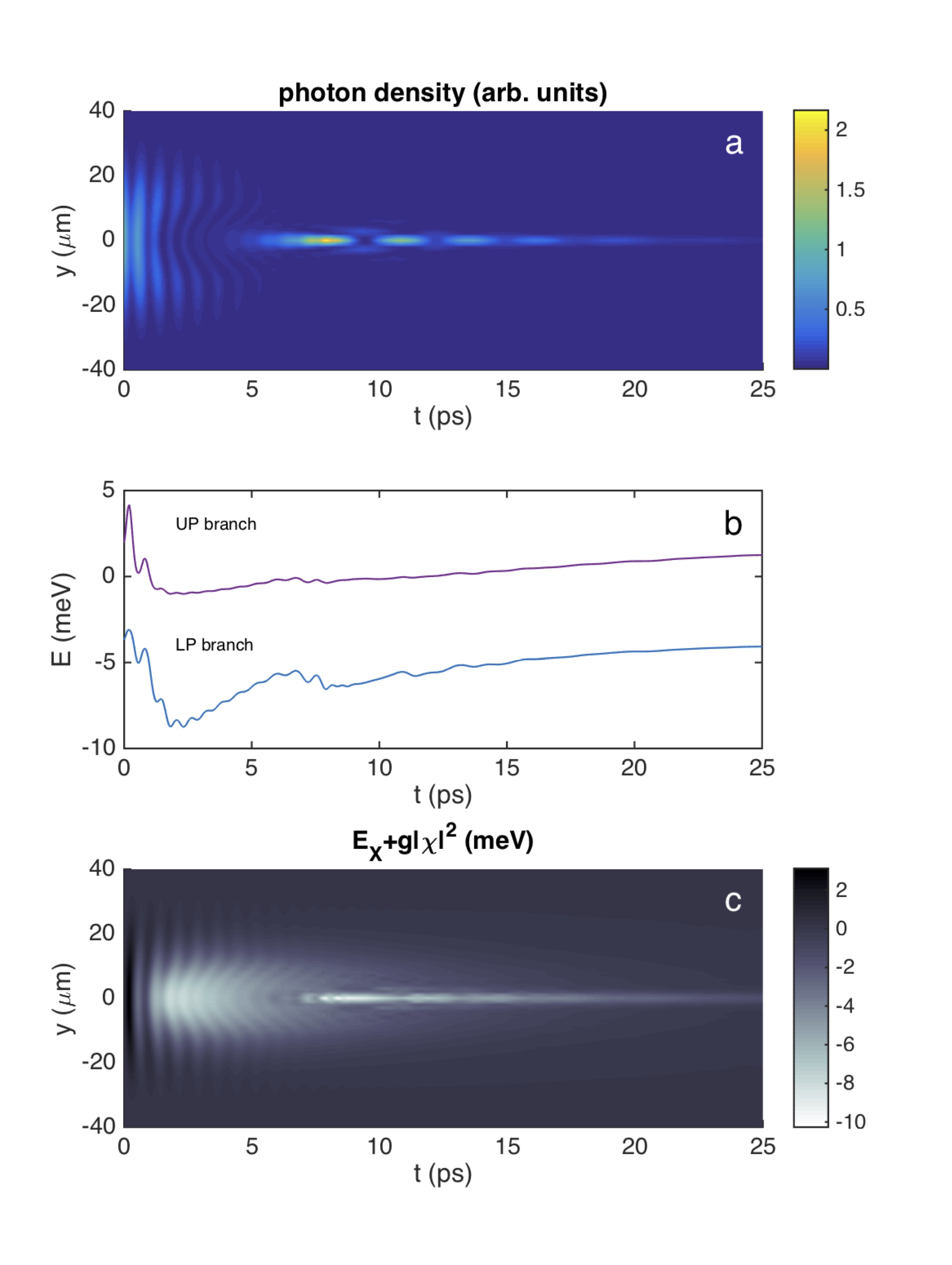}
\caption{(a) Time-space chart of the photonic fraction intensity
  distribution. (b) The time dependence of the local energies of the
  upper polariton and lower polariton branches. The time-space chart
  of the effective exciton potential is shown in panel (c). Parameters
  used in calculations are $\Omega_R = 2\protect\pi/
  (0.8\text{\;ps})$, $\Delta=-1$%
  ~meV, $m_C = 3\times10^{-5}$, $\protect\tau_C = 4$~ps,
  $\protect\tau_X = 8$%
  ~ps, $g = 2\cdot 10^{-2}$~meV \textmu{}m$^2$, $\protect\beta =
  g/n_\mathrm{%
    sat}$, $n_\mathrm{sat} = 3\times10^{10}$ cm$^{-1}$. The pumping
  field is described by a fs-pulse with $W=12.5$~\textmu{}m,
  $T_\mathrm{pulse} = 50$%
  ~fs, $E_p = 0$, and a variable amplitude $F_p$ (see text). }
\label{fig:Heating}
\end{figure}
The result of the numerical simulation of Eqs.~(\ref{eq:GPSystem})
with an Eq.~(\ref{eq:EX}) dependence is shown in
Fig.~\ref{fig:Heating}. At the initial stage, the Rabi oscillations
between photon and exciton states are seen. The dependence of the
oscillations frequency on the local detuning results in appearance of
the radial waves spreading from the center of the spot. The modified
density-dependent value of the Rabi frequency allows to reproduce some
retardation of their motion in the spot center. When the density
decreases due to the photonic decay, the direction of the
wave-spreading changes. In addition, the heating reservoir forms a
potential well, in which the polaritons are trapped through their
excitonic fraction, that results in a stabilization of both excitonic
and photonic components of the polariton flow. Further dynamics
exibits the decay of both components of the polariton many-body
wavefunction. One can see the typical behavior of the sharply
localised peak appearance extracted at different pumping power, which
is modeled by setting increasing amplitude of $F_{p}$ in
Eq.~\eqref{eq:GPSystem}. The results are shown in
Fig.~\ref{fig:DensityPowerDependence}. The photonic phase component is
plotted in Fig.~\ref{fig:PhasePowerDependence}. 
\begin{figure}[tb]
\includegraphics[width=.7\columnwidth]{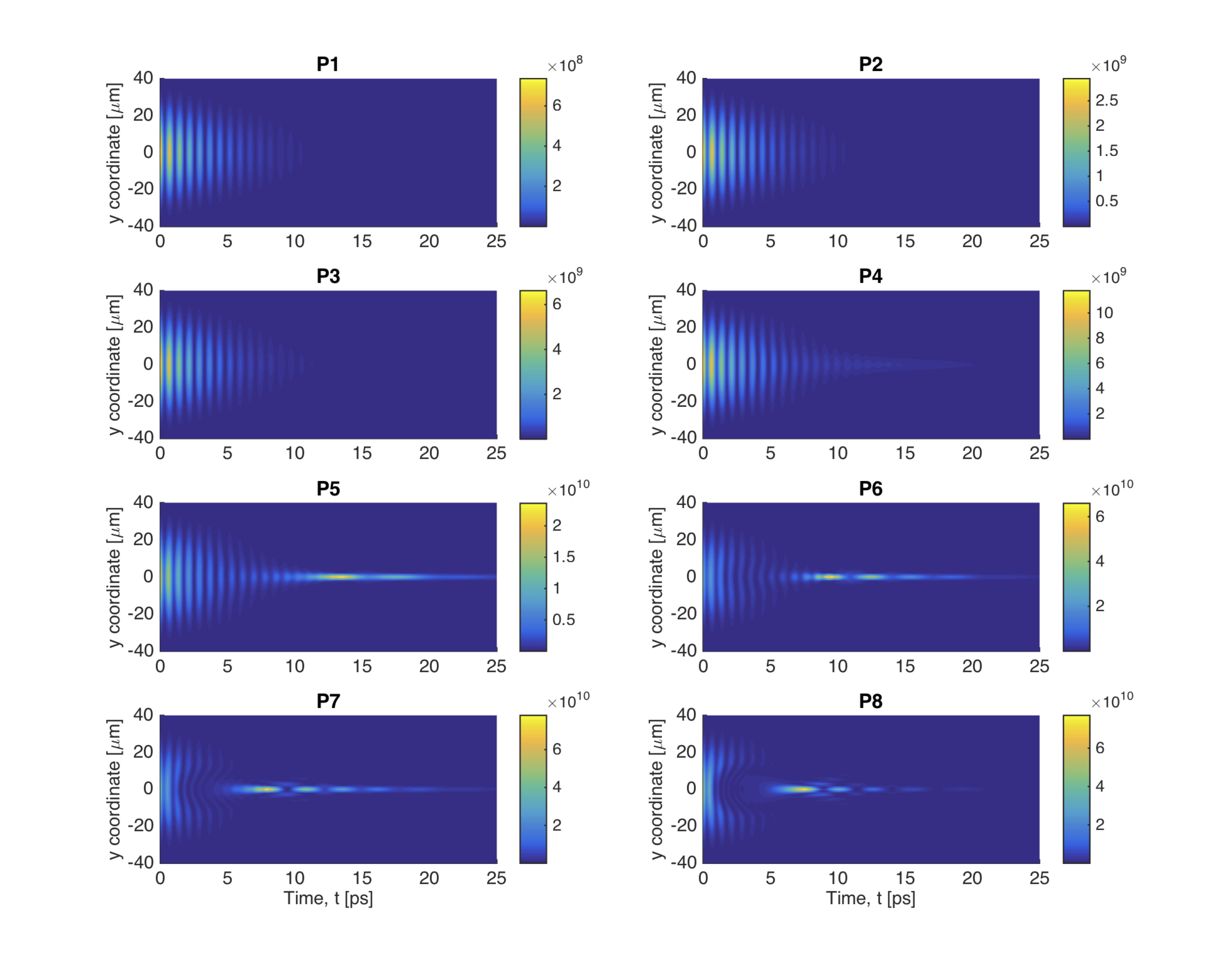}
\caption{Calculated time-space charts of the photonic component of the
  polariton wavefunction as in Fig.~\ref{fig:Heating}(a). The number
  of polaritons increases proportionally from $P_1$ to $P_8$. Other
  parameters are the same as in the caption of Fig.~\ref{fig:Heating}.}
\label{fig:DensityPowerDependence}
\end{figure}

\begin{figure}[bt]
\includegraphics[width=.7\columnwidth]{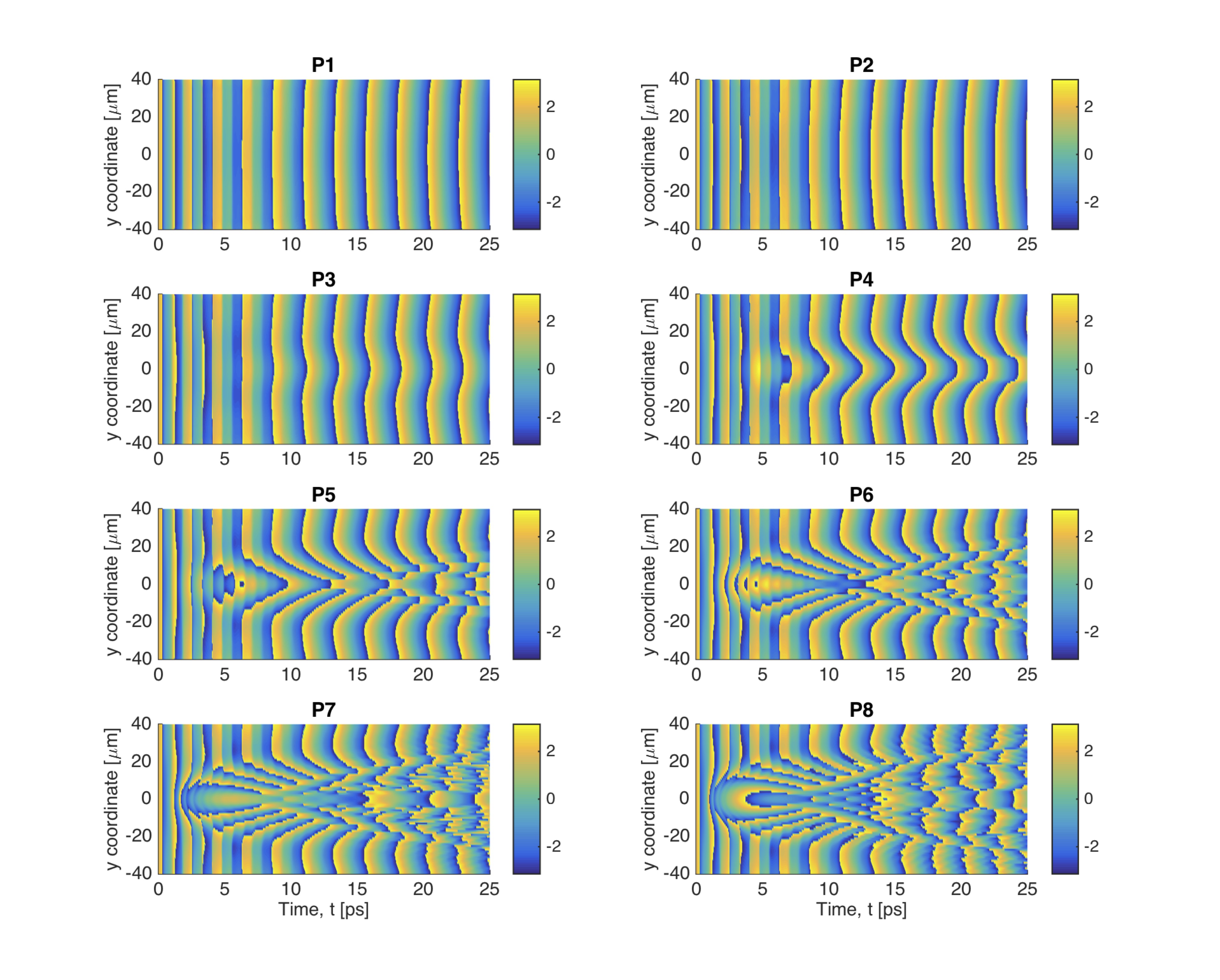}
\caption{Calculated time-space chart of the photonic phase at
  different powers if using a pulsed excitation, as in
  Fig.~\ref{fig:DensityPowerDependence}. }
\label{fig:PhasePowerDependence}
\end{figure}
In the present model as well, we assume that the incoherent reservoir
is either completely empty or does not play any role. This assumption
is certainly valid in the case of selective picosecond excitation of
the lower polariton (LP) branch. We note also that the weak-coupling
saturable optical nonlinearity could arise from instantaneous effects
or sample heating~\cite{S_Taranenko2005}.  The instantaneous part,
however, provides only defocusing nonlinearity for excitation below
the band edge, as follows from the nonlinear Kramers-Kronig
relations~\cite{S_Taranenko2005,S_Kost1998}. Heating generally leads to
redshift which corresponds to focusing nonlinearity but has a longer
response time.

In conclusions, while there is no compelling argument that impose the
collactive polaron effect as the mechanism causing the collapse of the
polariton condensate, our simulations show that it is a reasonable
candidate, when other more straightforward scenarios fail to provide
even a qualitative agreement or demand assumptions too strong to be
justified. It is not excluded that more exotic many-body physics is at
play, such as a variation of the BCS mechanism, holding the Bosonic
condensate together in a way similar as the superconducting phase
molds the Fermion liquid, or a dynamical Casimir effect~\cite{S_Koghee2014} 
pulling non-zero-momentum particles out of the suddenly hit polariton vacuum. 
We feel that it is left to theorists and
further investigations to elucidate which exact physics is causing the
peculiar phenomenology that we report.

\section*{Description of videos S1-S3}

\vskip-.1cm

Movies S1-S2. The movies show the complete imaging of the fluid dynamics over a
timespan of 40~ps with a timestep of 50~fs, as 3D and 2D maps,
respectively. This gives a direct view of the ultrafast superfluid
dynamics, combining Rabi oscillations and the central collapse along
with the other phenomena described in the text and in
Fig.~1-3.\newline 
Movie S3. Same as the previous movies but now
showing the corresponding radial cross section of the polariton
density $|\Psi(y)|^{2}$. As an interesting detail, it is possible to
observe how the initial expansion in the first few ps is happening
during the half-periods of the Rabi cycles where there is a low
photonic density (i.e., a high excitonic one).
\href{https://drive.google.com/folderview?id=0B0QCllnLqdyBZUNXMTdTNUNhdUk}{%
  Link to supplementary movies}


\begin{thebibliography}{46}
\bibitem{nota1}
\bibinfo{title}{Please note that the apparent discretization in energy is given
  by the interference in the energy domain of time-delayed reflections back
  from the substrate edge and is not affecting the ongoing dynamics; the
  substrate has got a 1.5 mm optical thickness, associated to a 10 ps time
  distance of the reflected echos of the emission and to a 0.45 mev fringes
  spacing in energy; the emission echo is visible also in the time-space
  measurements of real and reciprocal space, where it is seen weaker and weaker
  at regular interval of 10 ps.}
\end{thebibliography}

\end{document}